\documentclass[a4paper]{article}
\usepackage{latexsym,amssymb,amsmath}
\usepackage{amscd,array,cite,dsfont,enumitem,graphicx,mathtools,multirow,rotating,tensor,times,verbatim,youngtab}
\usepackage[bookmarksnumbered,bookmarksopen=true,bookmarksopenlevel=1,pdfstartview=FitH, colorlinks=true, linkcolor=blue, citecolor=black, urlcolor=blue]{hyperref}
\usepackage[all]{hypcap}
\usepackage[margin=2cm]{geometry}
\input xy
\xyoption{all}

\newcolumntype{M}[1]{>{$}{#1}<{$}}

\DeclareMathAlphabet\Scr{U}{rsf}{m}{n} \makeatletter
\@addtoreset{equation}{section} \makeatother

\newcommand{\be}{\begin{equation}}
\newcommand{\ee}{\end{equation}}
\newcommand{\bea}{\begin{eqnarray}}
\newcommand{\eea}{\end{eqnarray}}
\newcommand{\ba}{\begin{array}}
\newcommand{\ea}{\end{array}}
\newcommand{\bit}{\begin{itemize}}
\newcommand{\eit}{\end{itemize}}
\newcommand{\ben}{\begin{enumerate}}
\newcommand{\een}{\end{enumerate}}

\newcommand{\cq}{\mathcal{Q}}

\def\a{\alpha}

\allowdisplaybreaks[1]

\DeclareMathOperator{\sym}{Sym}

\DeclareMathOperator{\Aut}{Aut}
\DeclareMathOperator{\Str}{Str}
\DeclareMathOperator{\tr}{tr}
\DeclareMathOperator{\Tr}{Tr} % for partial traces only
\DeclareMathOperator{\Det}{Det}

\DeclareMathOperator{\diag}{diag}
\DeclareMathOperator{\Iso}{Iso}

\DeclareMathOperator{\SO}{SO}

\DeclareMathOperator{\SL}{SL}
\DeclareMathOperator{\SU}{SU}
\DeclareMathOperator{\Un}{U}
\DeclareMathOperator{\Sp}{Sp}

\newcommand{\FTS}{\mathfrak{F}}
\newcommand{\J}{\mathfrak{J}}

\newcommand{\F}{\mathds{F}}
\newcommand{\R}{\mathds{R}}
\newcommand{\C}{\mathds{C}}
\newcommand{\Q}{\mathds{H}}
\newcommand{\Oct}{\mathds{O}}
\newcommand{\da}{\mathds{A}}

\newcommand{\Z}{\mathds{Z}}
\newcommand{\N}{\mathcal{N}}

\newcommand{\half}{\ensuremath{\tfrac{1}{2}}}

\newcommand{\rep}[1]{\ensuremath{\mathbf{#1}}}

\newcommand{\ket}[1]{|#1\rangle}

\newtheorem{theorem}{Theorem}

\newtheorem{definition}[theorem]{Definition}

\begin{document}

\begin{titlepage}%1
\begin{center}
\hfill Imperial-TP-2020-MJD-02\\

\vskip 1.5cm

{\huge \bf Black Holes and Higher Composition Laws}

\vskip 1.5cm

{\bf L.~Borsten${}^a$, M.J.~Duff${}^b$, A. Marrani${}^c$}

\vskip 20pt

{\it ${}^a$Maxwell Institute and Department of Mathematics\\
Heriot-Watt University, Edinburgh EH14 4AS, United Kingdom\\\vskip 5pt
School of Theoretical Physics, Dublin Institute for Advanced Studies,\\
10 Burlington Road, Dublin 4, Ireland\\\vskip 5pt

${}^b$Institute for Quantum Science and Engineering and Hagler Institute for Advanced Study, Texas A\&M University, College Station, TX, 77840, USA\\\vskip 5pt
Theoretical Physics, Blackett Laboratory, Imperial College London,\\
 London SW7 2AZ, United Kingdom\\\vskip 5pt

${}^c$ Centro Ricerche Enrico Fermi, piazza del Viminale 1, I-00184 Roma, Italy \\\vskip 5pt
Dipartimento di Fisica e Astronomia `Galileo Galilei', Universit\'a di Padova, and
INFN, Sezione di Padova,
Via Marzolo 8, I-35131 Padova, Italy }\\\vskip 5pt
\href{mailto:l.borsten@hw.ac.uk}{\ttfamily l.borsten@hw.ac.uk}, ~\href{mailto:m.duff@imperial.ac.uk}{\ttfamily m.duff@imperial.ac.uk}, ~\href{mailto:alessio.marrani@pd.infn.it}{\ttfamily alessio.marrani@pd.infn.it}

\end{center}

\vskip 2.2cm

\begin{center} {\bf ABSTRACT}\\[3ex]\end{center}
We describe various relations between Bhargava's higher composition laws, which generalise Gauss's original composition law on integral binary quadratic forms,  and extremal  black hole  solutions appearing in string/M-theory and  related models. The cornerstone of these correspondences is the identification of the charge cube of the $STU$ black hole with Bhargava's cube of integers, which underpins the related higher composition laws.

%\today

%\end{center}

%\noindent

\vfill

%\July 2008

\end{titlepage}

\newpage \setcounter{page}{1} \numberwithin{equation}{section}
%\pagestyle{plain}
%\tableofcontents

\newpage\tableofcontents

\section{Introduction}

In 1801 Gauss introduced a beautiful
composition law on binary  quadratic forms with integer coefficients. In the modern parlance,  the set of $\SL(2, \Z)$-equivalence classes of primitive binary quadratic forms of a given discriminant $D$,  denoted $\text{Cl}(\text{Sym}^2(\Z^2)^*; D)$, has an inherent group structure, which  is isomorphic to  the narrow class group $\textrm{Cl}^{+}(S(D))$ of the unique quadratic ring $S(D)$ of discriminant $D$. This result is all the more remarkable  in light of the fact that groups were yet to be defined! This work was clearly  ahead of its time and,  in the words of Andrew Wiles, ``it came to a stop with Gauss''.  The original composition law  lay in waiting for a little over  200 years, when  Manjul Bhargava made  ground breaking progress by introducing a set of 14 (subsuming Gauss's original) \emph{higher} composition laws \cite{Bhargava:2004, bhargava2004higher, bhargava2004higherIII}.

In \cite{Moore:1998pn} Moore established a relationship between the arithmetics of supersymmetric black holes  in string theory and Gauss composition. Remarkably,  Bhargava's higher composition laws  are also closely related to various classes of black hole solutions appearing  in string/M-theory \cite{Borsten:2007rtn, Borsten:2008zz, Borsten:2008wd, Borsten:2009zy, Borsten:2010aa, Borsten:2010ths, Gunaydin:2019xxl}. In particular, Bhargava's higher law on $2 \times 2 \times 2$ hypermatrices, or \emph{cubes},  of integers is directly related to the extremal black hole solutions of the $STU$ model \cite{Sen:1995ff,Duff:1995sm}. By identifying \cite{Borsten:2007rtn, Borsten:2008zz, Borsten:2008wd, Borsten:2010ths} the black hole charge cube of \cite{Duff:1995sm} with the integers on the corners of Bhargava's cube, the U-duality equivalence classes of extremal $STU$ black hole charge vectors valued in $\Z^2\otimes\Z^2\otimes\Z^2$ with Bekenstein-Hawking entropy $S_{\text{BH}}=\pi\sqrt{|\Delta|}$ are seen to be in one-to-one correspondence  \cite{Borsten:2010ths} with pairs $(S, (I_1, I_2, I_3))$, where $S$ is the unique  quadratic ring with non-zero discriminant $D=-\Delta$ and $(I_1, I_2, I_3)$ is an equivalence class of balanced triples of oriented ideals, $I_1 I_2 I_3\subseteq S$ and $N(I_1)N(I_2)N(I_3)=1$. When restricting to \emph{projective}\footnote{Projectivity is the natural generalisation of primitivity for higher composition laws \cite{Bhargava:2004}.} $2 \times 2 \times 2$ cubes  Bhargava's composition law endows the equivalence classes with a group structure 
isomorphic to $\textrm{Cl}^{+}(S)\times\textrm{Cl}^{+}(S)$, where $\textrm{Cl}^{+}(S)$ denotes the narrow class group of the quadratic ring $S$ with discriminant $D=-\Delta$. Correspondingly, the U-duality equivalence classes of projective $STU$ black holes are characterised by $\textrm{Cl}^{+}(S)\times\textrm{Cl}^{+}(S)$ and  the corresponding class numbers count the physically distinct projective black hole  configurations  \cite{Borsten:2010ths}.

The relationship between Bhargava's composition  law on cubes and the black holes of the $STU$ model is but one example. It maintains a special status  as the  linchpin holding together all higher composition laws related to quadratic orders. Equivalently, the $STU$ model is the cornerstone of the various relevant (super)gravity theories. In the present contribution we describe the  relations between Bhargava's higher composition laws and black hole solutions in Einstein-Maxwell-scalar theories. Many of these correspond to supergravities with a string/M-theory origin, but supersymmetry is not essential. A number of the examples of have appeared in the literature before \cite{Borsten:2008zz, Borsten:2008wd, Borsten:2009zy, Borsten:2010aa, Borsten:2010ths, Gunaydin:2019xxl}.

The first six  examples of higher composition laws (including Gauss's original) introduced by Bhargava are obtained by
through various ``symmetrisations'' or embeddings of the  $2 \times 2\times 2$ cube  and are hence closely related to the $STU$ model. The new cases derived from the law on cubes are defined on: (1) pairs of binary quadratic forms; (2) binary cubic forms; (3) pairs of quaternary alternating 2-forms; and (4) senary  alternating 3-forms. In each case there is a corresponding (super)gravity theory. We describe in the detail these six cases as well as the additional cases of $E_{7(7)}(\Z)$ and $\SO(6,6; \Z)$ pertaining to $\N=8$ supergravity and its consistent truncation to $\SO(6,6; \Z)$ Maxwell-Einstein-scalar theory. In fact, all 14 of Bhargava's higher composition laws have a realisation in terms of black $p$-branes in (super)gravity theories, as remarked on in the conclusions. However, we shall return to the complete correspondence in future work. We also explain why such a correspondence between black hole charges and (super)gravity theories should exist at all.

\begin{table}\footnotesize
\[
\begin{array}{lllllllllllll}
\hline
\hline
&G(Z)&& V(\Z) && \text{Equivalence classes}&& \text{Projective  classes} && \text{(Super)gravity theory}\\ [5pt]
\hline

\\

1. &E_{7(7)}(\Z)&&  \FTS(\mathfrak{J}_{3}^{\mathfrak{O}_s})  && ?? && \{*\} && \mathcal{N}=8\\[5pt]

2. &\SO(6,6; \Z)&&  \FTS(\mathfrak{J}_{3}^{\mathfrak{H}_s})  && ?? && \{*\} && \mathcal{N}=0, 16A, 36\phi\\[5pt]

3. &\SL(6, \Z)&&  \wedge^3\Z^6\cong  \FTS(\mathfrak{J}_{3}^{\mathfrak{C}_s})   && S(D),  M_3 && \{*\} && \mathcal{N}=0, 10A, 20\phi\\[5pt]

4. &\SL(2, \Z)\times\SL(4, \Z)&& \Z^2\otimes \wedge^2\Z^4   && S(D), (I_S, M_2) && \textrm{Cl}^{+}(S(D)) && \mathcal{N}=0, 6A, 11\phi\\[5pt]

5. & \SL(2, \Z)\times\SL(2, \Z)\times\SL(2, \Z) && \Z^2\otimes\Z^2\otimes\Z^2&& S(D), (I_S, I_T, I_U) && \textrm{Cl}^{+}(S(D))\times\textrm{Cl}^{+}(S(D)) && \mathcal{N}=2, STU\\[5pt]
 
6. & \SL(2, \Z)\times\SL(2, \Z)&& \Z^2\otimes\text{Sym}^2(\Z^2) &&  S(D), (I_S, I_T, I_T) &&  \textrm{Cl}^{+}(S(D))  && \mathcal{N}=2, ST^2\\[5pt]

7. &\SL(2, \Z)&& \text{Sym}^3(\Z^2) && S(D), (I_T, I_T, I_T, \delta)&& \textrm{Cl}_3(S(D))&& \mathcal{N}=2, T^3 \\[5pt]

8. &\SL(2, \Z)&& \text{Sym}^2(\Z^2)^* && S(D), I&& \textrm{Cl}^+(S(D))&& \text{\cite{Moore:1998pn}}  \\[5pt]

\hline
\hline
\end{array}
\]
\caption{Summary of higher composition laws and black holes charge orbits in $D=4$. Note, case 8. admits an alternative interpretation as the black hole (or dual string) charge orbits of the (non-Jordan) $D=5$ $\N=2$ supergravity coupled to two vector multiplets with scalar manifold $\SO(1,2)/\SO(2)$ \cite{deWit:1992cr}, corresponding to case I.2 with $n=2$ in \S 7 of \cite{sato_kimura_1977}.}\label{quadratic}
\end{table}

\section{Black holes, higher composition laws and  quadratic orders}
In this section we shall describe the relationship between black holes and higher composition laws associated to ideal classes in quadratic orders. These are summarised in \autoref{quadratic}. Before treating the individual cases, we shall expand upon the notion of generalising Gauss composition in the context of quadratic orders and how/why it should be related to black holes appearing in ``gravity theories of type $E_7$''.

\subsection{Prehomogeneous vector spaces and group of type $E_7$}\label{prehomo}

What does it mean to generalise Gauss's composition law? One perspective,  advocated by Bhargava  \cite{bhargava2006higher} and also put to good use by Krutelevich \cite{Krutelevich:2004}, is motivated by the expression of Gauss's composition law as a parametrisation result:

\begin{theorem}\label{th1}
There is a canonical bijection between the set of $\SL(2, \Z)$-equivalence classes of nondegenerate binary quadratic forms $f(x,y)=ax^2+ b xy+ cy^2, ~a,b,c\in\Z$, and the set of isomorphism classes of pairs $(S , I )$, where $S$ is a nondegenerate oriented quadratic ring and $I$ is an oriented ideal class of $S$. 
\end{theorem}
Note, in the above we have not restricted to primitive forms, hence the lack of a group structure on the sets of equivalence classes. To summarise, we have a group $G(\Z)\cong\SL(2, \Z)$ with a representation $V(\Z)\cong\sym^2(\Z^2)^*$, such that the space of orbits $V(\Z)/G(\Z)$ is parametrised by ideal classes in a quadratic order. The two objects of \autoref{th1} are connected by the discriminant. On the one hand,  the unique algebraically independent  $\SL(2, \Z)$-invariant on binary quadratic forms is the discriminant, $D=b^2-4ac$, which take values $0, 1 \mod 4$. On the other, every quadratic order $S$ is determined up to isomorphism by its discriminant, $D=\text{Disc}(S)$, which also necessarily takes values  $0, 1 \mod 4$. The orbits of forms with discriminant  $D$ are one-to-one with ideal classes in $S$ with $D=\text{Disc}(S)$.

So, one way to interpret generalisations would be to seek $G$ and $V$ such that  $V(\Z)/G(\Z)$ is naturally parametrised by some algebraic structures. We will take ``naturally'' here to mean that the two sides of the picture are connected   by some $G$-invariant on the space $V$ that is bijectively mapped  to an  invariant characterising  the corresponding algebraic structure, for example,  the  discriminant in the  case of \autoref{th1}.

This shifts the question to one of identifying suitable candidate $G, V$. How should one go about this? Let us begin by noting that in the case of  \autoref{th1},  $\text{GL}(2, \C)$ has a single Zariski-open orbit\footnote{One might prefer to think of the family of $\SL(2, \C)$ orbits parametrised by the discriminant.} on the space of binary quadratic forms over $\C$ with non-vanishing  discriminant.  In view of \autoref{th1}, this can be thought of as following from the fact that over $\C$ there is a unique  class of pairs $(S, I)\cong\C\oplus\C$, where $S=I=\C\oplus\C$. Said another way, the pair $\text{GL}(2, \C), \sym^2(\Z^2)^*$ forms a prehomogeneous vector space:

\begin{definition}\label{prehom}
Let $G$ be  an algebraic group and $\rho$ a rational  representation on a vector space $V$. The triple $(G, \rho, V)$ is said to be a prehomogeneous vector space if the action of $G(\C)$ on $V(\C)$ has exactly one Zariski-dense orbit.
\end{definition}

Seeking  generalisations of \autoref{th1}, which are simple in the sense  that the algebraic side of the equation reduces to a single object when taken over $\C$, we should therefore focus on prehomogeneous vectors. Fortunately, the   reduced irreducible\footnote{See \cite{sato_kimura_1977} for definitions.}   prehomogeneous vector spaces have been classified, as presented in \S 7 of \cite{sato_kimura_1977}. If an irreducible prehomogeneous vector space is  regular\footnote{A prehomogeneous vector space is said to be regular if there is a relative $G$-invariant polynomial $f(x), x\in V$, with a not identically zero Hessian.}  it has a unique 
(up to a scalar factor) irreducible homogeneous relative  $G$-invariant polynomial  $f: V(\C)\rightarrow \C $ \cite{sato_kimura_1977}, which is the obvious candidate $G$-invariant underpinning our putative bijection to some algebraic structure.   For example, in the case of $(\text{GL}(2, \C), \rho, \sym^2(\C^2)^*)$ it is, of course, nothing but  the discriminant.  The six higher composition laws  related to quadratic orders, constructed  by Bhargava, are associated  to a special class of prehomogenous vector spaces for which $G(\C)$ is  of the form $G(\C)\cong\text{GL}(1, \C)\times G_7(\C)$ and $G_7$ is a group of type $E_7$ \cite{Brown:1969}.  

In fact, the relevant  $G_7$ appearing here are reduced groups of type $E_7$, which implies they are built on an underlying cubic Jordan algebra. In the seminal work of G\"unaydin, Sierra and Townsend \cite{Gunaydin:1983rk,Gunaydin:1983bi} it was shown that precisely such groups, with specific real forms, elegantly underpin the generic Jordan and magic $\N=2$ supergravity theories. It has since been understood, largely through the work of G\"unaydin, that these structures  appear  in many places in string/M-theory and black hole physics. See for example \cite{Gunaydin:1984ak, Gunaydin:1992zh, Ferrara:1997uz, Gunaydin:2000xr, Gunaydin:2005gd, Gunaydin:2005zz, Ferrara:2006xx, Bellucci:2006xz, Gunaydin:2005mx, Gunaydin:2007bg, Gunaydin:2007qq,Borsten:2008zz, Borsten:2008wd,Gunaydin:2009zza,Borsten:2009zy,Borsten:2010aa, Cerchiai:2010xv, Ferrara:2011gv, Ferrara:2011aa, Borsten:2011ai, Ferrara:2012qp, Cacciatori:2012sf, Borsten:2012pd, Borsten:2011nq,Cacciatori:2012cb, Marrani:2015wra, Chiodaroli:2015wal, Borsten:2013bp, Anastasiou:2013hba, Borsten:2017uoi, Borsten:2018djw}. In particular, see also \cite{Julia:1980gr, Cremmer:1999du} for early  examples and subsequent generalisations to magic triangles. The relationships between black holes in (super)gravity theories, prehomogeneous vector spaces and groups of type $E_7$ was introduced in \cite{Borsten:2009zy} and treated in detail in \cite{Marrani:2015wra}. Although the magic supergravities  are not themselves directly related to Bhargava's  higher composition laws (see comments in \autoref{comments}),  the appearance of  groups of type $E_7$ as the global symmetries  in various  (super)gravity theories is the first stone in the  bridge between black holes and higher composition laws.  

Accordingly, let us now briefly introduce groups of type $E_7$, which are  characterised as the automorphisms of  Freudenthal triple systems (FTS) \cite{Brown:1969}:

\begin{definition}\label{FTS} A FTS is a finite
dimensional vector space $\FTS$ over a field $\F$ (not of
characteristic 2 or 3), such that:
\begin{enumerate}
\item  $\FTS$ possesses a non-degenerate antisymmetric bilinear form $\{x, y\}.$
\item $\FTS$ possesses a symmetric four-linear form $q(x,y,z,w)$ which is not identically zero.
\item If the ternary product $T(x,y,z)$ is defined on $\FTS$ by $\{T(x,y,z), w\}=q(x, y, z, w)$, then
\be
3\{T(x, x, y), T(y,y,y)\}=\{x, y\}q(x, y, y, y).
\ee
\end{enumerate}
\end{definition}
For notational convenience, let us introduce 
\begin{subequations}\label{delta}
\begin{align} 
2\Delta(x, y, z, w)&\equiv q(x, y, z, w),  \\
\Delta(x)& \equiv\Delta(x, x, x, x),\\
T(x)& \equiv T(x, x, x).
\end{align}
\end{subequations}

\begin{definition}\label{autF} The \emph{automorphism} group of an FTS is defined  as the set of invertible  $\R$-linear transformations preserving the quartic and quadratic forms:
\be
\Aut(\FTS):=\{\sigma\in\Iso_\F(\FTS)|\{\sigma x, \sigma y\}=\{x, y\},\;\Delta(\sigma x)=\Delta(x)\},
\ee
\end{definition}
which implies $\sigma T(x,y,z)=T(\sigma x, \sigma y, \sigma z)$ and thus the automorphism group of the triple product. This defines   groups of type $E_7$. The prototypical example is, unsurprisingly,  given by  $E_7$, in which case $\FTS$ is the fundamental 56-dimensional representation and  the antisymmetric bilinear form and symmetric four-linear form are the unique symplectic quadratic and totally symmetric invariant in $\rep{56} \times_a \rep{56} $ and $\text{Sym}^4 (\rep{56})$, respectively. 

As shown in \cite{Brown:1969,Ferrar:1972}, every simple reduced\footnote{%
An FTS is simple if and only if $\{x ,y \}$ is non-degenerate,
which we assume. An FTS is said to be reduced if it contains a strictly
regular element: $\exists \;u\in \mathfrak{F}$ such that $T(u,u,u)=0$ and $%
u\in \text{ Range }L_{u,u}$ where $L_{x,y}:\mathfrak{F}\rightarrow \mathfrak{%
F};\quad L_{x,y}(z):=T(x,y,z)$. Note that FTS on \textquotedblleft
degenerate" groups of type $E_{7}$ (as defined in \cite{Ferrara:2012qp}, and
Refs. therein) are not reduced and hence cannot be written as $\mathfrak{F}(%
\mathfrak{J})$; they correspond to theories which cannot be uplifted to $D=5$
dimensions consistently reflecting the lack of an underlying $\mathfrak{J}$.}
 $\mathfrak{F}$ is isomorphic to an  $\mathfrak{F}(\mathfrak{J})$,
where
\begin{equation}\label{FFJJ}
\mathfrak{F}(\mathfrak{J}):=\mathds{F}\oplus \mathds{F}^*\oplus \mathfrak{J}%
\oplus \mathfrak{J}^*
\end{equation}%
and $\mathfrak{J}$ is the Jordan algebra of an admissible cubic form with
base point or the Jordan algebra of a non-degenerate quadratic form. With $\FTS$ regarded as an $\text{Aut}(\FTS)$-module, this corresponds to the decomposition of $\text{Aut}(\FTS(\mathfrak{J}))$ under $\text{Str}(\mathfrak{J})$, the structure group of $\mathfrak{J}$. The notation $\mathds{F}^*\cong\mathds{F}$   and  $\mathfrak{J}^*\cong\mathfrak{J}$ indicates that they  transform in conjugate representations of $\text{Str}(\mathfrak{J})$. For example, choosing  $\mathfrak{J}^{\mathds{O}}_3$, the Jordan algebra of $3\times 3$ Hermitian octonionic matrices,   we have $\text{Aut}(\FTS(\mathfrak{J}^{\mathds{O}}_3))\cong E_{7(-25)}$, $\text{Str}(\mathfrak{J}^{\mathds{O}}_3)\cong \Un(1)\times E_{6(-26)}$ and $\FTS(\mathfrak{J}^{\mathds{O}}_3)$ is the unique 56-dimensional $E_{7(-25)}$-module, which  decomposes as $\rep{56}\rightarrow\rep{1}_3+\rep{1}_{-3}+\rep{27}_1+\rep{{27}'}_{-1}$ under $\text{Str}(\mathfrak{J}^{\mathds{O}}_3)\subset \text{Aut}(\FTS(\mathfrak{J}^{\mathds{O}}_3))$. The FTS
quadratic form, quartic norm and triple product are then defined in terms of
the basic Jordan algebra operations \cite{Brown:1969}.

We shall write elements in the  basis \eqref{FFJJ} as 
\be
x=(\alpha, \beta, A, B),
\ee
 where $\alpha, \beta \in \F$ and $A, B\in \J$, respectively. Then 
 \begin{subequations}
    \begin{equation}\label{eq:bilinearform}
        \{x, y\}=\alpha\delta-\beta\gamma+\Tr(A,D)-\Tr(B,C),  
                \end{equation}
                for $x=(\alpha, \beta, A, B)$ and $y=(\gamma, \delta, C, D)$ and 
    \begin{equation}\label{eq:quarticnorm}
    \Delta (x)=-\left(\alpha\beta-\Tr(A,B)\right)^2-4[\alpha N(A)+\beta N(B)-\Tr(A^\sharp, B^\sharp)]
    \end{equation}
\begin{equation}\label{eq:Tofx}
T(x)=\left(-\alpha\kappa(x)-N(B), 
\beta\kappa(x)+N(A), 
-(\beta B^\sharp-B\times A^\sharp)+\kappa(x)A, 
(\alpha A^\sharp-A\times B^\sharp)-\kappa(x)B \right),
\end{equation}
\end{subequations} 
where $\Tr: \J\times \J^*\rightarrow \F$, $N: \J\rightarrow \F$ are the trace form and cubic norm of $\J$, respectively. For
details, see \cite{Borsten:2011nq} and references therein.  In \autoref{tab:FTSsummary} we list the relevant Jordan algebras, associated FTS and their automorphism groups.

We can then define an  \emph{integral}  $\FTS_\Z$, as introduced in the important work of \cite{Krutelevich:2004}, based on an integral Jordan algebra $\mathfrak{J}_\Z$ \cite{Krutelevich:2002},
\be\label{intFTS}
\FTS(\mathfrak{J}_\Z)\equiv \FTS_\Z\cong \Z\oplus\Z\oplus \mathfrak{J}_\Z%
\oplus \mathfrak{J}_\Z.
\ee
The integral structure on $\J_\Z$ implies that $N(A)\in\Z$ for all $A\in\J_\Z$, Hence,  from \eqref{eq:quarticnorm} we see that the quartic norm is quantised,
\be\label{quant}
\Delta (x)=0,1 \mod 4,\qquad \forall x\in  \FTS_\Z. 
\ee
That is, the allowed values of the quartic norm coincide precisely with those of the discriminant of a quadratic order. 

The automorphism group is broken to a discrete subgroup
\be
\Aut(\FTS_\Z):=\{\sigma\in\Iso_\Z(\FTS_\Z)|\{\sigma x, \sigma y\}=\{x, y\},\;\Delta(\sigma x)=\Delta(x)\}
\ee
 and is a model, in the sense of \cite{Gross:1996}, for $\Aut(\FTS)$ over $\Z$. In addition to the $\text{Aut}(\FTS)$-invariant quartic norm, there is a set of discrete invariants 
 \begin{equation}\label{Zinvariants}
    \begin{split}
    d_1(x)&=\gcd(x)\\
    d_2(x)&=\gcd(3\,T(x,x,y)+\{x,y\}\,x) \ \forall\ y\\
    d_3(x)&=\gcd( T(x) )\\
    d_4(x)&=|\Delta(x)|\\
    d'_4(x)&=\gcd(x \wedge T(x)),
    \end{split}
    \end{equation}
 where $\wedge$ denotes the antisymmetric tensor product. For reducible FTS we also have 
\be
     d'_2(x)=\gcd(B^\sharp -\alpha A,\,A^\sharp -\beta B,\, \mathcal{R}(x))
\ee
 where   $\mathcal{R}(x):\mathfrak{J}\to \mathfrak{J}$ is a Jordan algebra endomorphism given by
    \begin{equation}
    \mathcal{R}(x)(C)=2\left(\alpha\beta-\Tr(A,B)\right)C+2\{A,B,C\},\ C\in\mathfrak{J},
    \end{equation}
    where $\{A,B,C\}$ is the Jordan triple product.

 The integral Freudenthal triple systems, $\FTS_\Z$, and their corresponding  discrete automorphism groups, $\Aut(\FTS_\Z)$, will provide our  $V(\Z)$ and $G(\Z)$, respectively. By definition, they have the  desired property that,  when taken over $\C$, $\text{GL}(1, \C)\times G(\C)$ has one Zariski-dense orbit on  $V(\C)$. The quartic norm, $\Delta: \text{Sym}^4\FTS(\C)\rightarrow \C$,  provides the generalisation of the discriminant $D=b^2-4ac$; all elements in $V(\C)$ with non-vanishing  quartic norm   belong to the unique Zariski-dense orbit. Since over $\Z$ we have \eqref{quant}, the quartic norm can be consistently identified with the discriminant of a quadratic order $S(D)$, providing the link to the corresponding algebraic structures.

\begin{table}[tbp]
\caption[Jordan algebras, corresponding FTSs, and their associated symmetry
groups]{The automorphism group $\Aut(\mathfrak{F}(\mathfrak{J}))$ and the
dimension of its representation $\dim\mathfrak{F}(\mathfrak{J})$ given by
the Freudenthal construction defined over the cubic Jordan algebra $%
\mathfrak{J}$ over $\mathds{R}$ (with dimension $\dim\mathfrak{J}$ and
reduced structure group $\Str_0(\mathfrak{J})$).}
\label{tab:FTSsummary}%
\begin{tabular*}{\textwidth}{@{\extracolsep{\fill}}l*{5}{M{c}}l}
\hline
\hline
& \text{Jordan algebra }\mathfrak{J} & \Str_0(\mathfrak{J}) & \dim\mathfrak{J} & \Aut(\mathfrak{F}(\mathfrak{J})) & \dim\mathfrak{F}(\mathfrak{J}) &\\
\hline
1& \mathds{R}                     & -                                  & 1   & \SL(2,\mathds{R})                                   & 4    & \\
2& \mathds{R}\oplus\mathds{R}           & \SO(1,1)                       & 2   & \SL(2,\mathds{R})\times \SL(2,\mathds{R})                  & 6    & \\
3& \mathds{R}\oplus\mathds{R}\oplus\mathds{R} & \SO(1,1)\times \SO(1,1)    & 3   & \SL(2,\mathds{R})\times \SL(2,\mathds{R})\times \SL(2,\mathds{R}) & 8    & \\
4 & \mathds{R}\oplus \Gamma_{r,s}           & \SO(1,1)\times \SO(r,s)  & r+s+1 & \SL(2,\mathds{R})\times \SO(r+1, s+1)                & 2(r+s+2) & \\
5 & \mathfrak{J}_{3}^{\R}             & \SL(3, \mathds{R})                         & 6   & \Sp(6,\mathds{R})                                   & 14   & \\
6 & \mathfrak{J}_{3}^{\C}             & \SL(3,\mathds{C})                         & 9   & \SU(3,3)                                 & 20   & \\
6'& \mathfrak{J}_{3}^{\C^s}             & \SL(3,\mathds{R})\times\SL(3, \mathds{R})                         & 9   & \SL(6, \mathds{R})                                 & 20   & \\
7& \mathfrak{J}_{3}^{\Q}             & \SU^\star(6)                   & 15  & \SO^\star(12)                            & 32   & \\
7'& \mathfrak{J}_{3}^{\Q^s}             & \SL(6, \mathds{R})                   & 15  & \SO(6,6)                            & 32   & \\
8& \mathfrak{J}_{3}^{\Oct}             & E_{6(-26)}                   & 27  & E_{7(-25)}                            & 56   & \\
8'& \mathfrak{J}_{3}^{\Oct^s}             & E_{6(6)}                   & 27  & E_{7(7)}                            & 56   & \\
\hline
\hline
\end{tabular*}
\end{table}

\subsection{Einstein-Maxwell-Scalar theories of type $E_7$}

Following the preceding discussion it is perhaps unsurprising that when connecting to higher composition laws associated to quadratic orders, we consider black hole solutions in ``Einstein-Maxwell-scalar theories of type $E_7$'' or just  ``theories of type $E_7$'' for short. By theories of type $E_7$ we mean Einstein-Hilbert  gravity coupled to Abelian 1-forms $A$   and scalar fields $\phi$ (with no potential) such that: (i) the Gaillard-Zumino  electromagnetic duality group \cite{Gaillard:1981rj} is the automorphism group $\Aut(\FTS)$ of some $\FTS$; (ii) the Abelian field strengths together with  their duals take values in $\Lambda^2(M)\otimes\FTS$; (iii) and  the scalars parametrise the coset $\Aut(\FTS)/[\Aut(\FTS)]_\text{mcs}$, where $[G]_\text{mcs}$ denotes the maximal compact subgroup of $G$. They may or may not admit a (not necessarily unique) supersymmetric completion. Theories of type $E_7$ include all  $\mathcal{N}$-extended $D=1+3$ supergravities with $\mathcal{N}>2$ supersymmetries, as well as all $\mathcal{N}=2$ theories for which the scalar fields belonging to vector multiplets parametrise a symmetric space. Note,  however, for $\N=3$ supergravity (coupled to an arbitrary number of vector multiplets), as well as the minimally coupled $\N=2$ supergravities, the corresponding $\FTS$ are not reduced and their quartic invariant is  degenerate in the sense that it is the  square of a quadratic invariant. The `degeneration' of groups of type $E_7$ is discussed in \cite{Ferrara:2012qp}. The  best known example of a theory of type $E_7$ is provided the low energy effective field theory limit of type II string theory (M-theory) compactified on a 6-torus (7-torus). That is,  $D=1+3, \N=8$ supergravity, which has electromagnetic duality group $E_{7(7)}(\R)\cong \Aut(\mathfrak{J}_{3}^{\mathds{O}_s})$ that is broken to $E_{7(7)}(\Z) \cong \Aut(\mathfrak{J}_{3}^{\mathfrak{O}_s})$ by the Dirac-Zwanziger-Schwinger charge quantization condition. Here, $\Oct_s$ denotes the composition algebra of split octonions and ${\mathfrak{O}_s}$ the ring of integral split octonions. Another particularly  important example in the context of string/M-theory, is  type II string theory on $T^2\times K3$, which corresponds to example 4 of \autoref{tab:FTSsummary} with $r=5, s=21$. 

The two-derivative Einstein-Maxwell-scalar Lagrangian is uniquely determined by the choice of $\FTS$, although $\Aut(\FTS)$ is only a symmetry of the equations of motion due to electromagnetic duality; there is no conventional manifestly covariant  $\Aut(\FTS)$-invariant action for the field strengths. However, there does exist  a manifestly covariant $\Aut(\FTS)$-invariant Lagrangian \emph{if} one is willing to  accept  a twisted-self-duality constraint that must be imposed  on the Euler-Lagrange equations (but  not on the Lagrangian itself) \cite{Cremmer:1997ct, Borsten:2012pd}. The twisted-self-duality constraint implies both the Bianchi identities and equations of motion for the field strengths, so this construction would be redundant without the coupling to gravity/scalars always present in theories of type $E_7$. This formalism makes the notation compact and we adopt it here. Let us define the  ``doubled'' Abelian gauge potentials $ \mathcal{A}=(A, B)^T\in \Lambda^1(M)\otimes\FTS$ transforming as a symplectic vector  of $\Aut(\FTS)$, such that
  \be
 \mathcal{F}  =  d \begin{pmatrix} A \\ B \end{pmatrix},
 \ee
 and introduce the manifestly  $\Aut(\FTS)$-invariant Lagrangian,
  \be\label{doubleL}
 \mathcal{L}= R\star 1 +\frac{1}{4}\tr \left(\star d \mathcal{M}^{-1}\wedge d\mathcal{M}\right) -\frac{1}{4}  \star  \mathcal{F}  \wedge  \mathcal{M}   \mathcal{F},
 \ee
 with twisted-self-duality constraint \cite{Cremmer:1979up},
 \be\label{constraint}
 \mathcal{F}=\star \Omega \mathcal{M}    \mathcal{F},\qquad \Omega = -\mathds{1},\qquad \mathcal{M} \Omega\mathcal{M} ^{-1}=\Omega,
 \ee
 where $\mathcal{M}(\phi)$ is the $\Aut(\FTS)/[\Aut(\FTS)]_\text{mcs}$ scalar coset representative and $\mathcal{F}^T\Omega \mathcal{G}=\{ \mathcal{F},  \mathcal{G}\}$.
 The doubled Lagrangian \eqref{doubleL}, where the potential $\mathcal{A}$ is treated as the independent variable, together with  the constraint \eqref{constraint}, is on-shell equivalent to the standard Einstein-Maxwell-scalar Lagrangian  \cite{Cremmer:1997ct}.

For an Einstein-Maxwell-scalar theory of type $E_7$, the most general  extremal, asymptotically flat, spherically symmetric,
static,  dyonic black hole metric with non-vanishing horizon area is given by (cf.~for example \cite{Bellucci:2008sv} and the references therein)
\begin{equation}
ds^{2}=-e^{2U}dt^{2}+e^{-2U}(dr^{2}+r^{2}d\Omega _{2}),  \label{metric}
\end{equation}%
where $U=U(H\left( r\right))$ and
\be
e^{-2U}=\sqrt{|\Delta \left( H \right)|}, \quad 
H\left( r\right)=H_{\infty }-\frac{\mathcal{Q}}{r}.
\ee%
Here $H_{\infty }$ and $\mathcal{Q}$ belong to $\FTS$. The Abelian two-form fields strengths are given by 
\begin{equation}
\mathcal{F}  = \frac{e^{2U}}{r^{2}}\Omega \mathcal{%
M}\mathcal{Q}%
dt\wedge dr+\mathcal{Q}\sin \theta d\theta \wedge d\varphi, \label{vectors}
\end{equation}%
so that%
\begin{equation}
\frac{1}{4\pi }\int_{S_{\infty }^{2}}\mathcal{F}=\mathcal{Q}.\label{flux}
\end{equation}
The electromagnetic charges $\cq$ are elements of $\FTS$. Physically distinct charge configurations $\mathcal{Q}$ lie in distinct $\Aut(\FTS)$ orbits. The Bekenstein-Hawking entropy ($c=\hbar=G=1$) is given by 
\be
S_{\text{BH}} = \frac{A_{\text{hor}}}{4} = \pi \sqrt{|\Delta \left( \mathcal{Q} \right)|} =\pi \{\mathcal{Q}, \tilde{\mathcal{Q}}\}
\ee
where in the last equality we have used the Freudenthal dual $\tilde{x}:= \text{sgn}(\Delta(x)) T(x)/ \sqrt{|\Delta \left( x \right)|}$  \cite{Borsten:2009zy}. 

The  Dirac-Zwanziger-Schwinger charge quantisation conditions relating two black holes with charges $\cq$ and $\cq'$ are given by \cite{Borsten:2009zy}
 \be
 \{\cq, \cq'\}\in \Z.
 \ee
Consequently, the black hole charges belong to $\FTS_\Z$ as given in \eqref{intFTS}. The non-compact global symmetry group of \eqref{doubleL} is broken to $\Aut(\FTS_\Z)$, which corresponds to the U-duality group \cite{Hull:1994ys} in the context of M-theory. 

\subsection{Higher composition cube law and $STU$ Black Holes}

\subsubsection{Gauss composition and type II string theory on $T^2\times K3 $}
Before graduating to Bhargava's higher composition law on cubes and the $STU$ model, it will serve us well to first review  the relationship between Gauss composition and black hole solutions in type II string theory on $T^2\times K3$ appearing in Moore's  treatise on the arithmetic of black hole attractors  \cite{Moore:1998pn}.  The first $K3$ compactification was of $D=11$ supergravity on $T^3\times K3$ \cite{Duff:1983vj}, which yields precisely  the  massless sector of  type II supergravity on $T^2\times K3$ given by $D=4, \N=4$ supergravity coupled to 22 $\N=4$ vector multiplets. From the IIB perspective the black hole charges originating from the $D=10$ self-dual 5-form field strength belong to
\be
H^3(T^2\times K3 ; \Z)\cong H^1(T^2; \Z)\otimes H^2(K3 ; \Z) \cong II^{3, 19}\oplus II^{3, 19},
\ee
where $II^{3,19}$ is the even unimodular self-dual lattice of signature $(3,19)$. Similarly, the $D=10$ RR 2-form potential contributes $2+2$ charges. The $D=10$ graviton yields two $D=4$ Abelian gauge potentials, originating from the $T^2$ alone since the  $K3$ has no isometries, providing  $2+2$ charges. Finally, the NSNS 2-form  gives a further $2+2$ charges for a total of $28+28$ electromagnetic charges belonging to  $II^{6, 22}\oplus II^{6, 22}$.  The corresponding FTS in \autoref{tab:FTSsummary} is given by  case 4,   with $r=5, s=21$,   taken over $\Z$ , 
\be
\FTS_{T^2\times K3} \cong \Z\oplus\Z\oplus (\mathds{Z}\oplus \Gamma_{5,21}(\Z))\oplus (\mathds{Z}\oplus \Gamma_{5,21}(\Z)).
\ee
The Jordan algebra $\mathds{Z}\oplus \Gamma_{5,21}(\Z)$ is the space of electric (magnetic) $D=5$ black hole (string) charges   of type II on $S^1\times K3 $.  The U-duality group is given by 
\be
\text{Aut}(\FTS_{T^2\times K3}) \cong  \SL(2,\mathds{Z})\times \SO(6, 22; \Z).
\ee
The black hole charges, $\mathcal{Q}\in\FTS_{T^2\times K3 }$ transform in the $(\rep{2, 28})$ of  $\text{Aut}(\FTS_{T^2\times K3})$ and, in this case, can be written as
\be\label{senbasis}
\cq = \cq^{i}_{\mu}=\begin{pmatrix} P_\mu \\ Q_\mu \end{pmatrix}, \quad i =1,2 \quad \mu=1,2,\ldots 28. 
\ee
In this basis the quartic norm is given by 
\be
\Delta(\cq) = 4 \det \cq^{ij} = P^2Q^2 - (P\cdot Q)^2
\ee
where $\cq^{ij} = \cq^{i} \cdot \cq^{j}/2$ transforms as the $\rep{3}$ of $ \SL(2,\mathds{Z})$ and is manifestly  $ \SO(6, 22; \Z)$-invariant. 
To any $\cq^{i}_{\mu}\in \FTS_{T^2\times K3 }$ we can associate a binary quadratic form  via  $x_i=(x, y)$, 
\be
f_\cq(x, y) \equiv [a,b,c]_\cq = x_i \cq^{ij} x_j = a  x^2 + b yx + c y^2,
\ee
where $a =P^2/2, b=P\cdot Q, c=Q^2/2$. We observe that the discriminant is given by the quartic norm,   $D(f_\cq):=b^2 - 4ac=-\Delta(\cq)$.  Hence, we have a bijection between nondegenerate  binary quadratic forms $f_\cq$ and the triplets $\cq^{ij}$ derived from black hole charge configurations. Clearly, $f_\cq$ and $f_{\cq'}$ are   in the same equivalence class if and only if $\cq^{ij}$ and $\cq'^{ij}$ are $\SL(2, \Z)$ related. By \autoref{th1} the $\SL(2, \Z)$-equivalence classes $[\cq^{ij}]$   are in one-to-one correspondence with  the  isomorphism classes of pairs $(S , I )$, where $S$ is a nondegenerate oriented quadratic ring with discriminant $D=-\Delta(\cq)$ and $I$ is an oriented ideal class of $S$. Note,  all supersymmetric black hole solutions with non-vanishing entropy have $\Delta(\cq)>0$ and, hence, correspond to definite integral binary quadratic forms\footnote{The converse is not true as there are non-BPS solutions with $\Delta(\cq)>0$ \cite{Cerchiai:2009pi}. This is possible because $\SL(2,\mathds{R})\times \SO(6, 22)$ has two orbits for every $\Delta(\cq)>0$ \cite{Borsten:2011nq}, only one of which supports BPS black hole solutions \cite{Cerchiai:2009pi}.}. 

 As  observed in \cite{Moore:1998pn}, if $f_\cq$ is primitive (i.e.~$\gcd(a,b,c)=1$) then, by Gauss's theorem,  the set of $\SL(2, \Z)$-orbits of the associated primitive binary quadratic forms  having discriminant $D=-\Delta(\cq)$ naturally possesses the structure of a finite abelian group, the narrow class group $\text{Cl}^{+}(D)$ of the unique quadratic ring $S(D)$ of discriminant $D$.  The number, $n_D$, of  $\SL(2, \Z)$-orbits of  primitive $f_\cq$ with fixed  entropy $S_{\text{BH}}=\pi\sqrt{|\Delta(\cq)|}$  is then given by $|\text{Cl}^{+}(D)|$ \cite{Moore:1998pn}. Since for $f_\cq$ not primitive we can simply factor out $\gcd(f_\cq)$, it then follows \cite{Moore:1998pn} immediately that for arbitrary (not necessarily primitive)  $\cq^{ij}$ 
 \be
 n_D = \sum_{s}|\text{Cl}^{+}(D/s^2)|,\qquad \frac{D}{s^2}=0,1\mod 4.
 \ee
 Siegel \cite{siegel1935klassenzahl}  demonstrated that the number, $n_D$, of  $\SL(2, \Z)$-orbits of black holes    grows like $S_{\text{BH}}=\pi\sqrt{-D}$: $\forall \varepsilon>0, \exists c(\varepsilon)>0$ such that $n_D> c(\varepsilon) (S_{\text{BH}}){}^{1-2\varepsilon}$.

Of course,  if $\cq^{i}_{\mu}$ and $\cq'^{i}_{\mu}$ are $\SL(2,\mathds{Z})\times \SO(6, 22; \Z)$ related, then $\cq^{ij}$ and $\cq'^{ij}$ are $\SL(2,\mathds{Z})$-related. However, the converse is not necessarily true, contrary to the claims of \cite{Moore:1998pn}.  For $\SL(2,\mathds{Z})\times \SO(6, 22; \Z)$,  in addition to \eqref{Zinvariants}, which exist for any  $\FTS_\Z$, there is a further discrete invariant, the torsion \cite{Banerjee:2008ri},
\be
t(\cq) := \gcd(\varepsilon_{ij}\cq^{i}_{\mu}\cq^{j}_{\nu})=\gcd(P_{\mu}Q_{\nu}-P_{\nu}Q_{\mu}).
\ee
It is easy to find examples of pairs,  $\cq^{i}_{\mu}$ and $\cq'^{i}_{\mu}$, such that $\cq^{ij} = \cq'^{ij}$, but $t(\cq)\not=t(\cq')$. To illustrate this, we can restrict to an $\SL(2,\mathds{Z})\times \SO(2, 2; \Z)\subset \SL(2,\mathds{Z})\times \SO(6, 22; \Z)$ subsector, choosing primitive $\cq$ in the basis of \eqref{senbasis} given by 
\begin{equation}\label{stu_subsector}
P=\begin{pmatrix}Q_5\\J\\Q_1\\0\end{pmatrix},\quad Q =\begin{pmatrix}0\\n\\0\\1\end{pmatrix}, \qquad Q_1|J, Q_5
\end{equation}
with $\SO(2, 2; \Z)$-invariant metric
\be
\eta =\begin{pmatrix}0&\mathds{1}\\\mathds{1}&0\end{pmatrix}.
\ee
Here, $Q_5$ can be considered as representing an NS 5-brane wrapping charge, $n$ a fundamental string winding charge, while $J$ and $Q_1$ are units of KK monopole charge associated with the  two distinct circles of the $T^2$. In the canonical  FTS basis \eqref{FFJJ} we have
\begin{equation}\label{eq:sengeneral}
\cq =\begin{pmatrix}-1 & J &  (n, Q_5, Q_1) & (0, 0, 0) \end{pmatrix}. 
\end{equation}
Since, $Q_1|J, Q_5$ we can write 
\begin{equation}
P=Q_1 \begin{pmatrix}q_5\\j\\1\\0\end{pmatrix},\quad Q =\begin{pmatrix}0\\n\\0\\1\end{pmatrix}.
\end{equation}
Now let $Q_1=2m$ and consider a second primitive configuration  $\cq'$ given by
\begin{equation}
P'= m \begin{pmatrix}4q_5\\j\\1\\0\end{pmatrix},\quad Q' =\begin{pmatrix}0\\n\\0\\1\end{pmatrix}.
\end{equation}
Then $\cq^{ij} = \cq'^{ij}$, but $t(\cq)=2t(\cq')=2m$ so $\cq$ and $\cq'$ do not lie in the same U-duality orbit. This is our first indication that higher composition laws are relevant  to black holes; as we shall see,  the full U-duality orbits are related to the equivalence classes associated to higher composition laws. While the case of $\SL(2,\mathds{Z})\times \SO(6, 22; \Z)$ does not correspond to one of Bhargava's higher composition laws, it is closely related to the higher composition law on ``cubes'', which applies directly to the $STU$ model, as we shall explain. Viewed from this perspective, the connection to Gauss's original  composition law is a consequence of the fact that it is implied by the higher composition law on cubes.

\subsubsection{Bhargava's Cube Law and extremal $STU$ black holes}

In following we describe how the orbits of dyonic $STU$ black hole charges are characterised by Bhargava's higher composition law on ``cubes'' and ideal classes in quadratic orders \cite{Borsten:2010ths}. The $STU$ model, introduced independently in \cite{Sen:1995ff,Duff:1995sm}, provides an  interesting subsector of string compactifications to four dimensions. This model has a low energy limit which is described by $\N = 2$ supergravity coupled to three vector multiplets interacting through the special K\"ahler manifold  $[\SL(2, \R)/\SO(2)]^3$.   The three complex scalars are denoted by the letters $S, T$ and $U$, hence the name of the model \cite{Duff:1995sm,Behrndt:1996hu}. The remarkable feature that distinguishes it from generic $\N=2$ supergravities coupled to vectors \cite{Cremmer:1984hj} and, in particular, the $\N=2$ generic Jordan sequence \cite{Gunaydin:1984ak} given in case 4 of \autoref{tab:FTSsummary} with $r=1$, is its $S$-$T$-$U$ triality  \cite{Duff:1995sm}.  There are three different versions with two of the $\SL(2, \R)$ perturbative symmetries of the Lagrangian and the third a non-perturbative symmetry of the equations of motion. In a  fourth version all three are non-perturbative \cite{Duff:1995sm,Behrndt:1996hu}. All four are on-shell equivalent. If there are in addition four hypermultiplets, the $STU$ model is self-mirror \cite{Ferrara:2007pc, Duff:2010ss}. As a theory of type $E_7$ it is given by case 3  of \autoref{tab:FTSsummary} over $\Z$,
\be\label{STUFTS}
\mathfrak{F}_{STU}=\Z\oplus\Z\oplus\mathfrak{J}_{STU}\oplus\mathfrak{J}_{STU}
\ee
where $\mathfrak{J}_{STU}=\Z\oplus\Z\oplus\Z$. The  U-duality group\footnote{In the version of \cite{Sen:1995ff}, the discrete $\SL(2, \Z)$ are replaced by a subgroup denoted $\Gamma_0(2)$.} is given by 
\be
\Aut(\FTS_{STU})\cong \SL(2, \Z)\times\SL(2, \Z)\times\SL(2, \Z)\ltimes S_3,
\ee
 where the $S_3$ is the triality permutation group.  Equivalently, it is given by case 4 with $r=s=1$, which coincides with the subsector of type II string theory on $T^2\times K3$ used in \eqref{stu_subsector}.  More generally, the  $STU$ model can be considered as a consistent truncation of all theories of type $E_7$ with non-degenerate reduced FTS, with  the
exception of the  $ST^{2}$ and $T^{3}$  models, which are however obtained from  the $STU$ model by identifying $T=U$ and $S=T=U$, respectively. In this sense it is the key example. Similarly,  Bhargava's cubes provides the key example of a higher composition law.

The $STU$ black hole solutions have $4+4$ electromagnetic  charges, which in the canonical FTS basis \eqref{STUFTS}  are $\cq=(\alpha, \beta, (A_1, A_2, A_3), (B_1, B_2, B_3))$. See \S V. of  \cite{Borsten:2009zy}  for  details. In the physics literature they are typically split into the $4+4$ electric and magnetic charges  $\mathcal{Q}=(p^I, q_I), I=0,\ldots,3$, belonging to $\FTS_{STU}\cong \Z^2\otimes\Z^4$.  In \cite{Duff:1995sm} these charges were arranged into a cube and in \cite{Borsten:2008zz} this cube was identified with Bhargava's cube.  In \cite{Duff:2006uz} it was shown that the charges may also be arranged into  rank-three two-component tensor, or hypermatrix,  
\be
\ket{\mathcal{Q}}=a_{ABC}\ket{ABC}\in \Z^2\otimes\Z^2\otimes\Z^2,
\ee
where $A, B, C= 0, 1$, so that the the quartic norm is given  by Cayley's hyperdeterminant \cite{Cayley:1845}, making the triality symmetry manifest\footnote{This identification intiated what came to be known as the black-hole/qubit correspondence \cite{Duff:2006uz, Borsten:2008ur, Borsten:2008wd, Borsten:2010db, Borsten:2012fx}. Cayley's hyperdeterminant and the quartic norm, more generally, also give Nambu-Goto string actions \cite{Duff:2006ev, Nishino:2007ke, Borsten:2012pd}.}. That is  $\FTS_{STU}\cong \Z^2\otimes\Z^2\otimes\Z^2$. The three cases are trivially related by 
\be
\begin{split}
(\alpha, \beta, (A_1, A_2, A_3), (B_1, B_2, B_3)) &= (-q_0, p^0, (p^1, p^2, p^3), (q_1, q_2, q_3))\\
& = (-a_{111}, a_{000}, (-a_{001}, -a_{010}, -a_{100}), (a_{110}, a_{101}, a_{011})).
\end{split}
\ee
Bhargava also arranged the hypermatrix as a cube
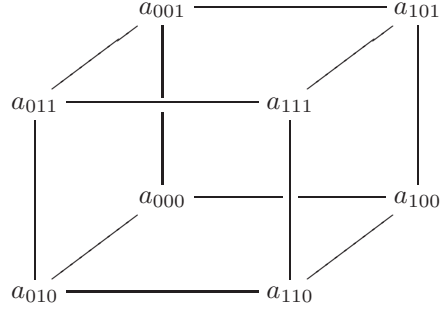
\[
\xymatrix{
& a_{001}  \ar@{-}[rr]\ar@{-}'[d][dd]
& & a_{101}  \ar@{-}[dd]
\\
a_{011} \ar@{-}[ur]\ar@{-}[rr]\ar@{-}[dd]
& & a_{111} \ar@{-}[ur]\ar@{-}[dd]
\\
& a_{000} \ar@{-}'[r][rr]
& & a_{100} 
\\
a_{010}\ar@{-}[rr]\ar@{-}[ur]
& & a_{110} \ar@{-}[ur]
}
\]
so that we speak of a higher composition law on $\SL(2, \Z)\times\SL(2, \Z)\times\SL(2, \Z)$-equivalence classes of cubes. 

In terms of the cube, the quartic norm is given by  Cayley's hyperdeterminant $\Det: \text{Sym}^4(\FTS_{STU})\rightarrow \Z$, where 
\begin{equation}\label{eq:CayleyHyperdeterminant}
\Delta(\mathcal{Q})= - \Det a =\half~\varepsilon^{A_1A_2}\varepsilon^{B_1B_2}\varepsilon^{A_3A_4}\varepsilon^{B_3B_4}\varepsilon^{C_1C_4}\varepsilon^{C_2C_3}   a_{A_1B_1C_1}a_{A_2B_2C_2}a_{A_3B_3C_3}a_{A_4B_4C_4}.
\end{equation}
Explicitly
\begin{equation}\label{eq:CayleyHyperdeterminant}
\begin{split}
\Det a =&\phantom{-\ }a_{000}^2 a_{111}^2 + a_{001}^2 a_{110}^2+a_{010}^2 a_{101}^2 + a_{100}^2 a_{011}^2 \\
&-2\,(\phantom{+\,}a_{000}a_{001}a_{110}a_{111}+a_{000}a_{010}a_{101}a_{111} \\
&\phantom{2\,(\ \ \ }+a_{000}a_{100}a_{011}a_{111}+a_{001}a_{010}a_{101}a_{110}\\
&\phantom{2\,(\ \ \ }+a_{001}a_{100}a_{011}a_{110}+a_{010}a_{100}a_{011}a_{101})\\ &+4\,(a_{000}a_{011}a_{101}a_{110}+a_{001}a_{010}a_{100}a_{111})
\end{split}
\end{equation}
Following \cite{Duff:2006ev} it is useful to  write
\begin{equation}\label{eq:q}
\Delta(\mathcal{Q})=4 \det\gamma^{S}= 4\det\gamma^{T}= 4\det\gamma^{U}=-\Det a,
\end{equation}
where we have defined the three matrices $\gamma^S,\gamma^T$, and $\gamma^U$
\begin{gather}
\begin{split}\label{eq:ABCgammas}
(\gamma^{S})_{A_{1}A_{2}}&=\half\varepsilon^{B_{1}B_{2}}\varepsilon^{C_{1}C_{2}}a_{A_{1}B_{1}C_{1}}a_{A_{2}B_{2}C_{2}}, \\
(\gamma^{T})_{B_{1}B_{2}}&=\half\varepsilon^{C_{1}C_{2}}\varepsilon^{A_{1}A_{2}}a_{A_{1}B_{1}C_{1}}a_{A_{2}B_{2}C_{2}}, \\
(\gamma^{U})_{C_{1}C_{2}}&=\half\varepsilon^{A_{1}A_{2}}\varepsilon^{B_{1}B_{2}}a_{A_{1}B_{1}C_{1}}a_{A_{2}B_{2}C_{2}},
\end{split}
\end{gather}
transforming respectively as $\rep{(3,1,1), (1,3,1), (1,1,3)}$ under $\SL(2, \Z) \times \SL(2, \Z) \times\SL(2, \Z) $. Explicitly,
\begin{gather}
\begin{split}
\gamma^{S}&=
\half\begin{pmatrix}
2(a_{0}a_{3}-a_{1}a_{2}) &  a_{0}a_{7}-a_{1}a_{6}+a_{4}a_{3}-a_{5}a_{2}\\
a_{0}a_{7}-a_{1}a_{6}+a_{4}a_{3}-a_{5}a_{2}  & 2(a_{4}a_{7}-a_{5}a_{6})
\end{pmatrix}, \\
\gamma^{T}&=
\half\begin{pmatrix}2(a_{0}a_{5}-a_{4}a_{1}) & a_{0}a_{7}-a_{4}a_{3}+a_{2}a_{5}-a_{6}a_{1}\\
a_{0}a_{7}-a_{4}a_{3}+a_{2}a_{5}-a_{6}a_{1} & 2(a_{2}a_{7}-a_{6}a_{3})
\end{pmatrix}, \\
\gamma^{U}&=
\half\begin{pmatrix}
2(a_{0}a_{6}-a_{2}a_{4}) &  a_{0}a_{7}-a_{2}a_{5}+a_{1}a_{6}-a_{3}a_{4}\\
a_{0}a_{7}-a_{2}a_{5}+a_{1}a_{6}-a_{3}a_{4} & 2(a_{1}a_{7}-a_{3}a_{5})
\end{pmatrix},
\end{split}
\end{gather}
where we have made a binary/decimal conversion $ABC\mapsto 2^0C+2^1B+2^2A$.
Bhargava's Cube Law provides a higher composition law for the $\SL(2, \Z)\times\SL(2, \Z)\times\SL(2, \Z)$-equivalences classes of \emph{projective} integer cubes $a_{ABC}$:
 \begin{definition}\label{projective}
A cube  $a_{ABC}$ is projective if the three  binary quadratic forms  defined by, 
\be\label{fstu}\begin{split}
f_S(x, y) = x^A (\gamma^{S})_{AA'} x^{A'} &= (a_{0}a_{3}-a_{1}a_{2})  x^2 +  (a_{0}a_{7}-a_{1}a_{6}+a_{4}a_{3}-a_{5}a_{2}) yx + (a_{4}a_{7}-a_{5}a_{6}) y^2,\\
f_T(x, y) = x^B (\gamma^{T})_{BB'} x^{B'} &= (a_{0}a_{5}-a_{4}a_{1})  x^2 +  (a_{0}a_{7}-a_{4}a_{3}+a_{2}a_{5}-a_{6}a_{1}) yx + (a_{2}a_{7}-a_{6}a_{3}) y^2,\\
f_U(x, y) = x^C (\gamma^{U})_{CC'} x^{C'} &= (a_{0}a_{6}-a_{2}a_{4})  x^2 +  (a_{0}a_{7}-a_{2}a_{5}+a_{1}a_{6}-a_{3}a_{4}) yx + (a_{1}a_{7}-a_{3}a_{5}) y^2,\\
\end{split}
\ee
where $x^A=(x, y)$, are each primitive.
\end{definition}
Note from \eqref{eq:q} that the discriminant of $f_S, f_T, f_U$ is given by the quartic norm $D(f_S)=D(f_T)=D(f_U)=\Det a$. As demonstrated by Bhargava, this construction provides an alternative definition of Gauss composition through the following theorem \cite{Bhargava:2004}:
 \begin{theorem}[Gauss composition from cubes]\label{cubegauss} The following statements hold and  are equivalent to Gauss composition: 
 
\begin{enumerate} 
\item Given a projective cube  $\ket{\mathcal{Q}}=a_{ABC}\ket{ABC}$  with quartic norm $\Delta(\cq)=-\Det a\not=0$, the Gauss composition of the associated forms $f_S, f_T, f_U$ is the identity element of  $\textrm{\emph{Cl}}^{+}(S(D))$, where $D=\Det a$. 
\item Given three primitive forms $f_1, f_2, f_3$ all of discriminant $D$ such that their Gauss composition is  the identity element of $\textrm{\emph{Cl}}^{+}(S(D))$, there exists a  cube  $a_{ABC}$ with hyperdeterminant  $\Det a=D$ such that $f_1=f_S, f_2=f_T, f_3=f_U$. 
\end{enumerate}
\end{theorem}
But Gauss composition concerns the set of $\SL(2, \Z)$-equivalence classes $\text{Cl}(\text{Sym}^2(\Z^2)^*; D)$ and, as we have already seen during the preceding discussion of type II strings on $K3$, it is the $\SL(2, \Z)\times\SL(2, \Z)\times\SL(2, \Z)$-equivalences classes that we need for the classification of the $STU$ black hole charge configurations \cite{Borsten:2009zy, Borsten:2010aa, Borsten:2010ths}. As one might anticipate by now, it is Bhargava's higher composition law that plays the analogous role. The $\SL(2, \Z)\times\SL(2, \Z)\times\SL(2, \Z)$-equivalences classes of projective cubes $a_{ABC}$ with discriminant $D=\Det a$ themselves form a group, which in analogy to the case of primitive binary forms Bhargava denoted $\text{Cl}(\Z^2\otimes\Z^2\otimes\Z^2; D)$. This is the first example of a higher composition law. It follows straightforwardly from \autoref{cubegauss}. Given two projective cubes $a$ and $a'$ we can define their composition $[a]+[a']=[a'']$, where $[x]$ denotes the $\SL(2, \Z)\times\SL(2, \Z)\times\SL(2, \Z)$-equivalence class of $x$, using the fact that  \autoref{cubegauss} implies $([f_S]+[f'_S]) +([f_T]+[f'_T])+([f_U]+[f'_U])=\text{Id}$ and  the existence and uniqueness, up to $\SL(2, \Z)\times\SL(2, \Z)\times\SL(2, \Z)$-equivalence, of a projective cube $a''$ such that $[f_S]+[f'_S]=[f''_S]$ and similarly for $T, U$. According as $D=0$ or $1$ mod 4, the identity element is given by the equivalence class of 
\be
\ket{\text{id}, D} =a^{\text{id}, D}_{ABC}\ket{ABC} = \ket{001}+\ket{010}+\ket{100}+\frac{D}{4}\ket{111}
\ee
or 
\be
\ket{\text{id}, D}=a^{\text{id}, D}_{ABC}\ket{ABC}  =  \ket{001}+\ket{010}+\ket{100}+\ket{110}+\ket{101}+\ket{011}+\frac{D+3}{4}\ket{111},
\ee
respectively. Note, in the context of three-qubit entanglement these are maximally entangled (unnormalised) GHZ states \cite{Borsten:2008ur, Borsten:2008wd, Borsten:2009yb, Borsten:2011is, Borsten:2013rca}.   In terms of the cube these correspond respectively to:
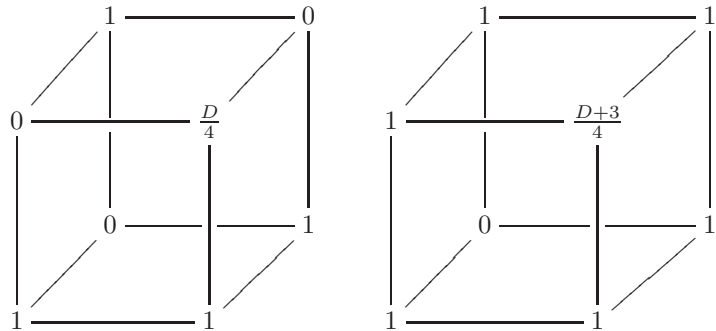
\[
\xymatrix{
& 1  \ar@{-}[rr]\ar@{-}'[d][dd]
& & 0 \ar@{-}[dd]
\\
0 \ar@{-}[ur]\ar@{-}[rr]\ar@{-}[dd]
& & \frac{D}{4} \ar@{-}[ur]\ar@{-}[dd]
\\
& 0 \ar@{-}'[r][rr]
& & 1
\\
1\ar@{-}[rr]\ar@{-}[ur]
& & 1 \ar@{-}[ur]
}
\qquad\xymatrix{
& 1  \ar@{-}[rr]\ar@{-}'[d][dd]
& & 1  \ar@{-}[dd]
\\
1 \ar@{-}[ur]\ar@{-}[rr]\ar@{-}[dd]
& & \frac{D+3}{4} \ar@{-}[ur]\ar@{-}[dd]
\\
&0 \ar@{-}'[r][rr]
& & 1
\\
1\ar@{-}[rr]\ar@{-}[ur]
& & 1\ar@{-}[ur]
}
\]

As for binary quadratic forms, using the perspective presented  in \autoref{prehomo}, we can rephrase this higher composition law as a special case of the following parametrisation result for generic (not necessarily projective) cubes in terms of ideal  classes in quadratic orders \cite{Bhargava:2004}:

\begin{theorem}\label{thcube}
There is a canonical bijection between the set of $\SL(2, \Z)\times\SL(2, \Z)\times\SL(2, \Z)$-equivalence classes of nondegenerate $(\Det a\not=0)$ cubes $a_{ABC}\ket{ABC}\in\Z^2\otimes\Z^2\otimes\Z^2$, and the set of isomorphism classes of pairs $S$ and  $[I_1, I_2, I_3 ]$, where $S$ is a nondegenerate oriented quadratic ring of discriminant $D=\Det a$ and $[I_1, I_2, I_3] $ is  an  equivalence class of a balanced triple of  ideals in $S$.
\end{theorem}
 The bijection is remarkably straightforward to state. To summarise,  given a black hole with charges $a_{ABC}$ the quadratic ring $S(D)$ is determined by the hyperdeterminant via $D = \Det a$, and the six bases for the ideal classes $I_i$ are determined by a system of eight equations involving $a_{ACB}$. Conversely, given any pair $S(D)$ and $[I_1, I_2, I_3]$  the corresponding black hole $a_{ABC}$ is directly obtained from the assumption that $I_1, I_2, I_3$ are a balanced triple.

In more detail, given $S(D)$, with oriented basis $\{1,\tau\}$ such that $\tau^2-\tfrac{D}{4}=0$ or $\tau^2-\tau+\tfrac{1-D}{4}=0$ according as $D= 0$ or $1\mod 4$, and a balanced triple $(I_S, I_T, I_U)$ with bases $\alpha_A, \beta_B, \gamma_C$, respectively, we have by assumption
\be\label{abc}
\alpha_A\beta_B\gamma_C  = c_{ABC}+ a_{ABC}\tau.
\ee
The claim is that $a_{ABC}$ are the black hole charges with $S_{\text{BH}}=\pi\sqrt{|\Det a |}$ under the bijection. If $(I_S, I_T, I_U)$ is replaced by an equivalent triple, $a_{ABC}$ is left invariant. An orientation preserving basis change on each ideal induces an $\SL(2, \Z)\times\SL(2, \Z)\times \SL(2, \Z)$ transformation on $a_{ABC}$.  So,  the equivalence classes  of pairs $S(D), (I_S, I_T, I_U)$ are mapped injectively  into the $\SL(2, \Z)\times\SL(2, \Z)\times \SL(2, \Z)$-orbits in $\Z^2\otimes\Z^2\otimes\Z^2$ through \eqref{abc}. Conversely, $a_{ABC}$ determines $[S(D), (I_S, I_T, I_U)]$ uniquely. First,  the balanced hypothesis implies $\Det a=D$ for \eqref{abc}, fixing $S(D)$. Second, the balanced assumption  implies that the $c_{ABC}$ are determined  uniquely by $a_{ABC}$ through,
\be\label{STU-T}
c_{ABC} =- \frac{1}{2}\left(T(a)_{ABC} + (\Det a\mod 4) a_{ABC}\right), 
\ee
which is in $\Z^2\otimes\Z^2\otimes\Z^2$. Here, $T$ is the triple product \eqref{eq:Tofx} of the associated FTS. The triple product was not used in \cite{Bhargava:2004}, but its appearance here makes the connection between Freudenthal triple systems, groups of type $E_7$ and higher composition laws manifest. For the $STU$ model \eqref{eq:Tofx}  is given by the three identical expressions
\begin{equation}
\begin{split}
T_{A_3B_1C_1}=-\varepsilon^{A_1A_2}a_{A_1B_1C_1}(\gamma^{A})_{A_{2}A_{3}}\\
T_{A_1B_3C_1}=-\varepsilon^{B_1B_2}a_{A_1B_1C_1}(\gamma^{B})_{B_{2}B_{3}}\\
T_{A_1B_1C_3}=-\varepsilon^{C_1C_2}a_{A_1B_1C_1}(\gamma^{C})_{C_{2}C_{3}}.
\end{split}
\end{equation}
where explicitily 
\begin{equation}
\begin{split}
T_0&=a_0 \left(\phantom{-}a_3 a_4+a_2 a_5+a_1 a_6-a_0a_7\right)-2 a_1 a_2 a_4 \\
T_1&=a_1 \left(          -a_3 a_4-a_2 a_5+a_1a_6-a_0 a_7\right)+2 a_0 a_3 a_5 \\
T_2&=a_2 \left(          -a_3 a_4+a_2a_5-a_1 a_6-a_0 a_7\right)+2 a_0 a_3 a_6 \\
T_3&=a_3 \left(          -a_3a_4+a_2 a_5+a_1 a_6+a_0 a_7\right)-2 a_1 a_2 a_7 \\
T_4&=a_4 \left(\phantom{-}a_3a_4-a_2 a_5-a_1 a_6-a_0 a_7\right)+2 a_0 a_5 a_6 \\
T_5&=a_5 \left(\phantom{-}a_3 a_4-a_2a_5+a_1 a_6+a_0 a_7\right)-2 a_1 a_4 a_7 \\
T_6&=a_6 \left(\phantom{-}a_3 a_4+a_2 a_5-a_1a_6+a_0 a_7\right)-2 a_2 a_4 a_7 \\
T_7&=a_7 \left(          -a_3 a_4-a_2 a_5-a_1 a_6+a_0a_7\right) +2 a_3 a_5 a_6.
\end{split}
\end{equation}
 This emphasises the implicit role played by the FTS in the higher composition laws. A balanced triple yielding $c_{ABC}, a_{ABC}$ exists and is determined uniquely up to equivalence by $c_{ABC}, a_{ABC}$ establishing the bijection.

The key point is that the physically distinct extremal $STU$ black charge configurations can be characterised entirely in terms of ideal classes in quadratic orders.  The higher composition law on cubes of fixed $D$ is then straightforwardly stated through the product of equivalence classes of balanced triples defined by,
\be
[S(D); I_S, I_T, I_U] \circ [S(D); I'_S, I'_T, I'_U] := [S(D); I_SI'_S, I_TI'_T, I_UI'_U]
\ee

If we restrict to the set of projective elements  $\text{Cl}(\Z^2\otimes\Z^2\otimes\Z^2; D)$, then under the bijection we restrict  to invertible ideals and $I_U$ is determined by $I_S, I_T$.   Consequently, the set of isomorphisms classes has a group structure given by the product of two copies of the narrow class group
\be
\text{Cl}(\Z^2\otimes\Z^2\otimes\Z^2; D)\cong\text{Cl}^+(S(D))\times \text{Cl}^+(S(D)),
\ee 
and \autoref{cubegauss} induces a group homomorphism
\be
\text{Cl}(\Z^2\otimes\Z^2\otimes\Z^2; D)\rightarrow \text{Cl}(\text{Sym}^2(\Z^2)^*; D).
\ee 

Hence, the physically distinct $STU$ extremal large black holes charge configurations $\cq$ with a fixed Bekenstein-Hawking entropy $S_{\text{BH}}=\pi\sqrt{|\Delta(\cq)|}$ are given by the isomorphism classes of pairs of quadratic orders of discriminant $D=-\Delta(\cq)$ and  balanced triples  of fractional ideals \cite{Borsten:2010ths}. If we restrict to projective black holes then the set of physically distinct configurations of a fixed entropy $S_{\text{BH}}=\pi\sqrt{|\Delta(a)|}$ has a group structure isomorphic to $\text{Cl}^+(S(D))\times \text{Cl}^+(S(D))$, where $D=-\Delta(\cq), $ and number  
 of distinct physically distinct projective configurations,
 \be 
n_{D}^{\text{proj}}:=|\left\{[\cq] | \Delta(\cq)=-D,\; \gcd(f_S)=\gcd(f_T)=\gcd(f_U)=1\right\}|,  
\ee
is given by $|\text{Cl}^+(S(D))|^2$ \cite{Borsten:2010ths}.  Interestingly, the number of U-duality inequivalent projective $STU$ black holes is a square number. Note, projectivity implies primitivity, $\gcd(\cq)=1$, but the converse is not true. For example, in the FTS basis 
\be
(1,0, (1, 2, 2), (0,0,0))
\ee
is primitive, but not projective as $f_S=[-2,0,-2]$. Consequently, one cannot generically divide through by  $\gcd(\cq)$ to make $\cq$ projective and the straightforward counting of equivalence classes of generic binary quadratic forms in terms of class numbers for primitive forms does not work for generic cubes.

\subsubsection{$STU$ U-duality orbits and quadratic orders: examples}

Let us make these ideas more concrete with some simple examples. It is straightforward to show that every black hole charge configuration can be brought into a five parameter canonical form using U-duality \cite{Borsten:2009zy}, which in the FTS basis may be expressed as 
\be
\cq_{\text{can}} = \alpha (1, j, (a, b, c), (0, 0, 0)). 
\ee
The corresponding forms (for $\alpha=1$) are given by 
\be\begin{split}
f_S(\cq_{\text{can}}) &= [-ab, -j, c ],\\
f_T(\cq_{\text{can}}) &= [-ca  , -j ,b ],\\
f_U(\cq_{\text{can}})  &= [-bc , -j ,a ].\\
\end{split}
\ee
The discrete invariants \eqref{Zinvariants} and three torsions $t_S, t_T, t_U$ on $\cq_{\text{can}}$ are given by 
 \begin{equation}
    \begin{split}
    d_1(\cq_{\text{can}})&=\alpha\\
    d_2(\cq_{\text{can}})&=\alpha^2 \gcd(j, 2a, 2 b, 2 c)\\
        t_S(\cq_{\text{can}})&=\alpha^2 \gcd(j, a,  b)\\
                t_T(\cq_{\text{can}})&=\alpha^2 \gcd(j, a,  c)\\
        t_U(\cq_{\text{can}})&=\alpha^2 \gcd(j,  b,  c)\\
    d_3(\cq_{\text{can}})&=\alpha^3 \gcd(j, 2 b c, 2 a c , 2 a b)\\
    d_4(\cq_{\text{can}})&=\alpha^4|j^2+4abc|\\
    d'_4(\cq_{\text{can}})&=2\alpha^4 \gcd( j^2+a b c, b c, a c, a b).
    \end{split}
    \end{equation}
Projectivity implies $\alpha=1$ and 
 \begin{equation}
    \begin{split}
 \gcd(a, j, bc)&=1\\
 \gcd(b, j, ca)&=1\\
 \gcd(c, j, ab)&=1\\
    \end{split}
    \end{equation}
Note, projectivity implies primitivity and that torsion in the $S, T$ and $U$ frame is one, i.e.~configurations with non-trivial torsion are not projective. 

Let us consider some simple examples of projective  black holes with $D<0$, which includes all supersymmetric configurations. For $D<0$ we have $\text{Cl}^+(S(D))\cong\text{Cl}(S(D))$, the class group. Hence,
 \be 
n_{D<0}^{\text{proj}}=|\text{Cl}(S(D))|^2. 
\ee
Consequently,   we see that  $n_{D<0}^{\text{proj}}$ grows like $S_{\text{BH}}^2$ as $D\rightarrow\infty$.

 As a first simple special case take those  black hole configurations of fixed Bekenstein-Hawking entropy on which U-duality acts transitively, that is $\Delta(\cq)=D$ for $D$ such that $|\text{Cl}(S(D))|=1$. This amounts to the classic Gauss class number problem for class number one, solved by Heegner \cite{Heegner1952}, Baker \cite{baker_1966} and  Stark \cite{stark1967}. For even $D=-4n,  n\in \mathds{N}$,  $|\text{Cl}(S(D))|=1$ iff $n=1,2,3,4,7$ (this answers Gauss's original question). We can parametrise these orbits by letting $j=0, \alpha=a=b=1$ and $c=-n$, 
  \be
  \begin{split}
 \gcd(1, 0, n)&=1,\\
 \gcd(1, 0, n)&=1,\\
 \gcd(n, 0, 1)&=1,\\
    \end{split}
    \end{equation}
so that every projective supersymmetric black hole with $\Delta(\cq)>0$ even is U-duality equivalent to one of  five canonical forms, 
\be
\cq_{\text{BPS, even}} =  (1, 0, (1, 1, -n), (0, 0, 0)), \quad n=1,2,3,4,7,
\ee
where $n$ is determined uniquely by  the U-duality invariant $\Delta(\cq_{\text{BPS, even}}) = 4n$. Regarded as  embedded in type II string theory on $T^2\times K3$, every such black hole is U-duality equivalent to  
one  wrapped NS 5-brane, one  wrapped fundamental string and    $n$  units of KK monopole charge associated with the second  circle of the $T^2$. 

The odd discriminants $D<0$ with class number one, are 
\be
-3, -7, -11, -19, -27, -43, -67, -163. 
\ee
A projective black hole has $\Delta$ odd iff $j$ is odd (an odd number of KK monopole charge).   Let  us just turn on just one unit of KK monopole charge $j=1$ on the first circle of $T^2$, then 
\be
\cq_{\text{BPS, 1, even}} =  (1, 1, (1, 1, -n), (0, 0, 0)),   
\ee
yields
\be
\Delta(\cq_{\text{BPS, even}}) = -(1-4n)=3, 7, 11, 19, 27, 43, 67, 163,
\ee
for a prime  $n=1, 2, 3, 5, 7, 11, 17, 41$ units of KK charge, respectively, as required (for what it is worth $n$, excluding $n=1$, is given by the first seven primes which are not the sums of two consecutive non-Fibonacci numbers). This exhausts all cases where U-duality acts transitively on the set of projective black holes of a fixed entropy.  

Let us consider some examples from the next simplest case, $\text{Cl}(D)\cong \Z_2$ corresponding to four U-duality orbits. Again, at class number two we have an exhaustive list of possible $D<0$. The simplest example, is given by $D=-15$ \cite{ireland1990classical}. Since there are four orbits, we are seeking four $\cq$ such that $\Delta(\cq)=15$ with distinct U-duality invariants, but are projective and hence torsionless. Representatives of the four classes are given by,
\be
\begin{split}
\cq_1&=  (1,  -1, (-1, 1, 4), (0, 0, 0)),\\
\cq_2&=  (1,  -1, (-1, 2, 2), (0, 0, 0)),\\
\cq_1&=  (1,  -1, (1, 4, -1), (0, 0, 0)),\\
\cq_2&=  (1,  -1, (2, -1, 2), (0, 0, 0)).\\
\end{split}
\ee
The corresponding forms are given by,
\be
\begin{array}{lllllll}
&&f_S && f_T && f_U\\[5pt]
\cq_1 && [1,1,4] 	&& [4,1,1] 	&& [-4,1,-1]\\[5pt]
\cq_2 && [2,1,2] 	&& [2,1,2]		&& [-4,1,-1]\\[5pt]
\cq_3 && [-4,1,-1] 	&&[1,1,4]		&&[4,1,1] \\[5pt]
\cq_4 && [2,1,2] 	&&[-4,1,-1]	&&[2,1,2] 
\end{array}
\ee
Using elementary $\SL(2, \Z)$ operations we can bring them all into reduced form
\be
\begin{array}{lllllll}
&&f_S && f_T && f_U\\[5pt]
\cq_1 && [1,1,4] 	&& [1,1,4] 	&& [-1,-1,-4]\\[5pt]
\cq_2 && [2,1,2] 	&& [2,1,2]		&& [-1,-1,-4]\\[5pt]
\cq_3 && [-1,-1,-4] 	&&[1,1,4]		&&[1,1,4] \\[5pt]
\cq_4 && [2,1,2] 	&&[-1,-1,-4]	&&[2,1,2] 
\end{array}
\ee
from which we immediately see that none are $\SL(2, \Z)\times \SL(2, \Z)\times\SL(2, \Z)$-equivalent, although $\cq_1$ $(\cq_2)$ and $\cq_3$ $(\cq_4)$ are triality related.

\subsection{Symmetrising the Cube law: further examples}
In the following we consider the four remaining examples of Bhargava's higher composition laws related to quadratic orders. These are related to the composition law of cubes discussed in the previous section by symmetrisation and embeddings. In particular, the symmetrisations yield examples relevant to   the $\N=2$ $ST^2$ and $T^3$ supergravity models, which admit a stringy derivation. The embeddings give two Einstein-Maxwell-Scalar theories of type $E_7$ that are $\N=0$ consistent truncations of $\N=8$ supergravity. Finally, we reconsider the analysis of $\mathcal{N}=8$ supergravity given in \cite{Borsten:2010aa}, which although not related to a non-trivial composition law, can be treated on an equal footing following \cite{Krutelevich:2004}. 
\subsubsection{The $ST^2$ model}

The $ST^2$ model is  $\N=2$ supergravity coupled to two vector multiplets with scalars belonging to 
\be
\frac{\SL_S(2, \R)\times \SL_{T^2}(2, \R)}{\SO(2)\times \SO(2)}.
\ee 
As a theory of type $E_7$, it corresponds to case 2 of \autoref{tab:FTSsummary} over $\Z$, 
\be
\FTS_{ST^2}\cong \Z \oplus\Z\oplus\J_{ST^2}\oplus\J_{ST^2},
\ee
 where $\J_{ST^2}\cong \Z\oplus\Z$. 
It  can be regarded as a ``symmetrisation'' of the $STU$ model, where the complex structure and K\"ahler forms are identified $T=U$, along with the two vector multiplets they sit in. It can be uplifted to pure $\N=(1,0)$ minimal chiral supergravity in $D=6$.  

The black hole charges  are symmetrised, $\cq=a_{A(BB')} \ket{A(BB')} \in\FTS_{ST^2} \cong \Z^2\otimes\sym^2 (\Z^2)$, corresponding to the natural inclusion 
\be
\imath_{ST^2}: \Z^2\otimes\sym^2 (\Z^2)\hookrightarrow \Z^2\otimes\Z^2\otimes\Z^2: a_{A(BB')} \mapsto a_{ABB'} 
\ee
where $a_{A(BB')}\in\Z$ and $a_{A(BB')}=a_{A(B'B)}$.
 This corresponds to the doubly-symmetric cube:
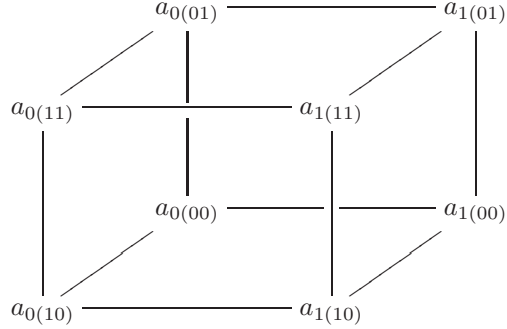
\[
\xymatrix{
& a_{0(01)}  \ar@{-}[rr]\ar@{-}'[d][dd]
& & a_{1(01)}  \ar@{-}[dd]
\\
a_{0(11)} \ar@{-}[ur]\ar@{-}[rr]\ar@{-}[dd]
& & a_{1(11)} \ar@{-}[ur]\ar@{-}[dd]
\\
& a_{0(00)} \ar@{-}'[r][rr]
& & a_{1(00)} 
\\
a_{0(10)}\ar@{-}[rr]\ar@{-}[ur]
& & a_{1(10)} \ar@{-}[ur]
}
\]
  Correspondingly, the U-duality group is given by the S-duality $\SL_S(2, \Z)$ together with the diagonal subgroup $\SL_{T^2}(2, \R)\subset \SL_T(2, \R)\times \SL_{U}(2, \R)$.

One can regard $\Z^2\otimes\sym^2 (\Z^2)$ as the space of doublets of (classical,  in the Gauss sense) integral binary quadratic  forms,
\be 
f_{A}(x,y) = a_{A(BB')}x^B x^{B'} = \begin{pmatrix} a_{000} x^2+  2a_{0(01)}xy +a_{011}y^2 \\ a_{100} x^2+  2a_{1(01)}xy +a_{111}y^2\end{pmatrix}.
\ee

This identification gives us a parametrisation of U-duality equivalence classes of  black hole charge configurations in terms of ideal classes in quadratic orders:

 \begin{theorem}[Parametrisation of equivalence classes of pairs of classical integral binary quadratic forms \cite{Bhargava:2004}]
There is a canonical bijection between the set of nondegenerate $\SL(2, \Z) \times \SL(2, \Z)$-orbits on the space $\Z^2\otimes\sym^2 (\Z^2)$, and the set of isomorphism classes of pairs $S(D), (I_S, I_T, I_U)$, where $S(D)$ is a nondegenerate oriented quadratic ring and $(I_S, I_T, I_U)$ is an equivalence class of balanced triples of oriented ideals of $S(D)$ such that  $I_T=I_U$. 
\end{theorem}
Note,  the discriminant is defined by the quartic norm $D=-\Delta(\cq)$.  The higher composition law on doubly-symmetrised cubes of fixed $D$ is then straightforwardly stated through the product of equivalence classes of balanced triples defined by,
\be
[S(D); I_S, I_T, I_T] \circ [S(D); I'_S, I'_T, I'_T] := [S(D); I_SI'_S, I_TI'_T, I_TI'_T].
\ee

This bijection follows the same logic as the equivalent statement for the $STU$ model. We will not repeat it here, other than to note that the solution to the required system of equations
\be\label{acc}
\alpha_A\beta_B\beta_{B' } = c_{A(BB')}+ a_{A(BB')}\tau.
\ee
is again given by the triple product of the associated FTS,
\be
c_{A(BB')} =- \frac{1}{2}\left(T(a)_{A(BB')} + (\Det a\mod 4) a_{A(BB')}\right).
\ee

An element $\cq\in \Z^2\otimes\sym^2 (\Z^2)$  is defined to be projective if the associated cube $\imath_{ST^2}(\cq)$ is projective. Under this inclusion 
\be
f^{\text{id}, D}_{A}(x,y) = \begin{pmatrix}   2xy  \\  x^2 +\frac{D}{4}y^2\end{pmatrix},\qquad f^{\text{id}, D}_{A}(x,y) = \begin{pmatrix} 2xy +y^2 \\ x^2+  2xy +\frac{D+3}{4}y^2\end{pmatrix}
\ee
map to the identity cubes in $\Z^2\otimes\Z^2\otimes\Z^2$ for $D=0,1$ mod 4, respectively. 
This identification then provides a higher composition law with a group structure on projective pairs of classical binary integral forms:

 \begin{theorem}[Group law on projective pairs of classical binary quadratic  forms \cite{Bhargava:2004}]\label{hcl24} For all $D=0,1 \mod 4$,  the set of ~$\SL(2,\Z)\times\SL(2, \Z)$-equivalence classes of projective doublets of  binary quadratic forms is a unique group $\text{\emph{Cl}}(\Z^2\otimes\sym^2 (\Z^2) ;D)$ such that

\begin{enumerate}[label=\alph*)]
\item $[f^{\text{\emph{id}}}_{A}]$ is the additive identity
\item The inclusion $\imath_{ST^2}: \Z^2\otimes\sym^2 (\Z^2)\hookrightarrow \Z^2\otimes\Z^2\otimes\Z^2$ induces a group homomorphism 
\be
\begin{array}{cccccccc}
\tilde{\imath}_{ST^2}:&\text{\emph{Cl}}(\Z^2\otimes\sym^2 (\Z^2) ;D)&\rightarrow&\text{\emph{Cl}}(\Z^2\otimes\Z^2\otimes\Z^2 ;D)\\
			&	[f_A]&\mapsto &\tilde{\imath}_{ST^2}([f_A]):=[\imath_{ST^2}(f_A)]
\end{array}
\ee
\end{enumerate}
\end{theorem}
Computing the binary quadratic forms corresponding to $\imath_{ST^2}(\cq)$ given in \eqref{fstu},
\be\begin{split}
f_S(x, y)&= (a_{0}a_{3}-a_{1}^{2})  x^2 +  (a_{0}a_{7}+a_{4}a_{3}-2a_{5}a_{1}) yx + (a_{4}a_{7}-a_{5}^{2}) y^2,\\
f_T(x, y)&= (a_{0}a_{5}-a_{4}a_{1})  x^2 +  (a_{0}a_{7}-a_{4}a_{3}) yx + (a_{1}a_{7}-a_{5}a_{3}) y^2,\\
f_U(x, y)&= (a_{0}a_{5}-a_{1}a_{4})  x^2 +  (a_{0}a_{7}-a_{3}a_{4}) yx + (a_{1}a_{7}-a_{3}a_{5}) y^2,\\
\end{split}
\ee
we observe that $f_T=f_U$. If projective,  any two of $f_S, f_T, f_U$ determines the third and $f_T=f_U$, hence the map taking $a_{A(BB')}$ to $f_T$ induces an isomorphism $\text{{Cl}}(\Z^2\otimes\sym^2 (\Z^2) ;D)\cong \text{{Cl}}(\sym^2 (\Z^2)^* ;D)\cong\text{Cl}^+(S(D))$. The U-duality equivalence classes of the projective black holes with entropy $S_{\text{BH}}=\pi \sqrt{|\Delta(\cq)|}$ of the $ST^2$ model are characterised precisely by the narrow class group $\text{Cl}^+(S(D))$, where $D=-\Delta(\cq)$.

\subsubsection{The $T^3$ model}

The $T^3$ model is  $\N=2$ supergravity coupled to a single vector multiplet with complex scalar belonging to 
\be
\frac{ \SL(2, \R)}{\SO(2)}.
\ee 
It follows from the $S^1$ dimensional reduction of pure $\N=2$ supergravity in $D=5$. As a theory of type $E_7$, it corresponds to case 1 of \autoref{tab:FTSsummary} over $\Z$, 
\be
\FTS_{T^3}\cong \Z \oplus\Z\oplus\J_{T^3}\oplus\J_{T^3},
\ee 
where $\J_{T^3}\cong \Z$. 

It  can be regarded as a further ``symmetrisation'' of the $ST^2$ model, with the axion-dilaton and   complex structure   identified $S=T$, along with the two vector multiplets they sit in. Consequently the black hole charges  are symmetrised, $\cq=a_{(AA'A'')}\ket{(AA'A'')}\in\FTS_{T^3} \cong \sym^3 (\Z^2)$, corresponding to the natural inclusion 
\be
\imath_{T^3}: \sym^3 (\Z^2) \hookrightarrow \Z^2\otimes\Z^2\otimes\Z^2: a_{(AA'A'')} \mapsto a_{AA'A''} 
\ee
where $a_{(AA'A'')}\in\Z$ and $a_{(AA'A'')}=a_{(\sigma(A)\sigma(A')\sigma(A''))}$, where $\sigma\in S_3$.  This corresponds to the triply-symmetric cube:
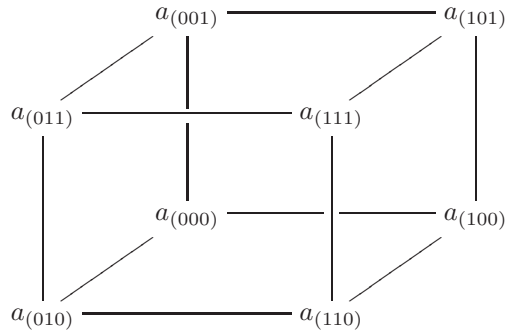
\[
\xymatrix{
& a_{(001)}  \ar@{-}[rr]\ar@{-}'[d][dd]
& & a_{(101)}  \ar@{-}[dd]
\\
a_{(011)} \ar@{-}[ur]\ar@{-}[rr]\ar@{-}[dd]
& & a_{(111)} \ar@{-}[ur]\ar@{-}[dd]
\\
& a_{(000)} \ar@{-}'[r][rr]
& & a_{(100)} 
\\
a_{(010)}\ar@{-}[rr]\ar@{-}[ur]
& & a_{(110)} \ar@{-}[ur]
}
\]
Correspondingly, the U-duality group is given by the  diagonal subgroup $\SL_{T^3}(2, \R)\subset \SL_S(2, \R)\times \SL_{T^2}(2, \R)$.

One can regard $\sym^3 (\Z^2)$ as the space of classical integral binary cubic  forms,
\be 
f^{\cq}(x,y) = a_{(AA'A'')}x^Ax^{A'} x^{A''} = a_{(000)} x^3+  3a_{(001)}x^2y +3a_{(011)}xy^2 +a_{(111)}y^3.
\ee
As before, the discriminant of $f$ is  related to  quartic norm by $D(f)=-\Delta(\cq)$. 

 \begin{theorem}[Parametrisation of equivalence classes of classical integral binary cubic forms \cite{Bhargava:2004}]
There is a canonical bijection between the set of nondegenerate $\SL(2, \Z)$-orbits on the space $\sym^3(\Z^2)$ of binary cubic forms of discriminant $D$, and the set of equivalence classes of triples $(S(D), I, \delta)$, where $S$ is a nondegenerate oriented quadratic ring with discriminant $D$, $I$ is an ideal of $S$, and $\delta$ is an invertible element of $S \otimes \mathds{Q}$ such that $I^3\subset \delta S$ and $N(I)^3 = N(\delta)$. 
\end{theorem}
 The higher composition law on triply-symmetrised cubes of fixed $D$ is then inherited   from the law on cubes,
\be\label{t3prod}
[S(D); I, I, I] \circ [S(D); I', I', I'] := [S(D); II', II', II'].
\ee
Of course, this can be mapped to $[S(D); I] \circ [S(D); I'] := [S(D); II']$, which superficially looks like Gauss composition. However, the notation in \eqref{t3prod} serves to remind us that the ideals $I$ are balanced,  $I^3\subset \delta S$ and $N(I)^3 = N(\delta)$, so they are indeed distinct, albeit closely related, composition laws.  Again,  the solution to the required system of equations underpinning the bijection 
\be\label{ccc}
\alpha_A\alpha_{A'}\alpha_{A'' } = c_{(AA'A'')}+ a_{(AA'A'')}\tau.
\ee
is  given by the triple product of the associated FTS,
\be
c_{(AA'A'')} =- \frac{1}{2}\left(T(a)_{(AA'A'')} + (\Det a\mod 4) a_{(AA'A'')}\right),
\ee
which is the symmetrisation of \eqref{STU-T}. 
 The cubic $f^{\cq}$ is defined to be projective if the associated cube $\imath_{T^3}(\cq)$ is projective. Note, this is not the same as the cubic itself being primitive.  Explicitly, for \eqref{fstu} evaluated on $\imath_{T^3}(\cq)$ we have 
 \be
 f_S=f_T=f_U= (a_{0}a_{3}-a_{1}^{2})  x^2 +  (a_{0}a_{7}-a_{3}a_{1}) yx + (a_{1}a_{7}-a_{3}^{2}) y^2,
 \ee
 so $f^{\cq}$ is projective if $\gcd(a_{0}a_{3}-a_{1}^{2}, a_{0}a_{7}-a_{3}a_{1}, a_{1}a_{7}-a_{3}^{2})=1$. 
  Under this inclusion 
\be
f^{\cq_{\text{id}, D}}(x,y) =   3x^2y  +\frac{D}{4}y^3,\qquad f^{\cq_{\text{id}, D}}(x,y) = 3x^2y  + 3xy^2 +\frac{D+3}{4}y^3
\ee
map to the identity cubes in $\Z^2\otimes\Z^2\otimes\Z^2$ for $D=0,1$ mod 4, respectively. 
This identification then provides a higher composition law with a group structure on projective pairs of classical binary integral forms:

 \begin{theorem}[Group law on projective   classical binary cubic  forms \cite{Bhargava:2004}]\label{hcl4} For all $D=0,1 \mod 4$,  the set of ~$\SL(2,\Z)$-equivalence classes of projective    binary cubic forms of discriminant $D$ forms a unique group $\text{\emph{Cl}}(\sym^3 (\Z^2) ;D)$ such that

\begin{enumerate}[label=\alph*)]
\item $[f^{\cq_{\text{id}, D}}]$ is the additive identity
\item The inclusion $\imath_{T^3}: \sym^3 (\Z^2)\hookrightarrow \Z^2\otimes\Z^2\otimes\Z^2$ induces a group homomorphism 
\be
\begin{array}{cccccccc}
\tilde{\imath}_{T^3}:&\text{\emph{Cl}}(\sym^3 (\Z^2) ;D)&\rightarrow&\text{\emph{Cl}}(\Z^2\otimes\Z^2\otimes\Z^2 ;D)\\
			&	[f^\cq]&\mapsto &\tilde{\imath}_{T^3}([f^\cq]):=[\imath_{T^3}(f^\cq)]
\end{array}
\ee
\end{enumerate}
\end{theorem}
This corresponds to the surjective group  homomorphism  
\be
\text{{Cl}}(\sym^3 (\Z^2)  ;D)\rightarrow \{g\in \text{Cl}(S(D))| g^3=\text{id}\}.
\ee
This yields the curious result that the number of physically distinct projective black hole solutions with fixed entropy $\pi\sqrt{|\Delta(\cq)|}$ corresponds to the number of invertible ideal classes  in $S(D)$, where $D=-\Delta(\cq)$, having order three in $\text{Cl}(S(D))$. 

\subsubsection{The $\SL(2, \Z)\times \SL(4, \Z)$ Einstein-Maxwell-Scalar theory}
The theory of type $E_7$ given by case 4 of \autoref{tab:FTSsummary} with $r=2, s=2$ has quantised electromagnetic duality group 
\be
\Aut(\FTS(\R\oplus\Gamma_{1,5}))  \cong  \SL(2, \Z)\times \SO(3, 3; \Z) \cong  \SL(2, \Z)\times \SL(4, \Z).
\ee
It corresponds to Einstein-Hilbert gravity coupled to six Abelian gauge fields, whose field strengths and duals  belong to the $\rep{(2,6)}$ of $\SL(2, \R)\times \SO(3, 3; \R)$, and eleven  real scalar fields parametrising the coset,
\be
\frac{\SL(2, \R)\times \SO(3, 3)}{\SO(2)\times \SO(3)\times\SO(3)}.
\ee
This theory  has been treated previously in \cite{Marrani:2017aqc, Marrani:2019jvd}. Although it does not admit a supersymmetric completion\footnote{Since all  $\N\leq 4$ multiplets have an even number of scalars.} in $D=1+3$, it can be regarded as a consistent truncation of $\N=8$ supergravity effected through the branching under 
\be
\begin{split}
E_{7(7)} & \supset  \SL(2, \R)\times \SO(6, 6) \supset\SL(2, \R)\times [ \SO(1,1)\times \SL(6, \R)] \\
&\supset \SL(2, \R)\times [ \SO(1,1)\times (\SL(2, \R)\times\SL(4, \R))] 
\end{split} 
\ee
so that the $28+28$ vectors  and their duals break as 
\be
\rep{56}\rightarrow \rep{(2,12)+(1,32)}\rightarrow\rep{20} \rightarrow \rep{(2,6)}
\ee
where we retain only the $\SL(2,\R)$ and $\SO(1,1)$ singlets at the first and second branchings, respectively.  From the above we see that the $STU$ may be embedded  in the $\SL(2, \Z)\times \SL(4, \Z)$ Einstein-Maxwell-scalar theory, further branching under $\SO(1,1)\times \SL(2, \R)\times \SL(2, \R)\subset\SL(4, \R)$ and retaining only the $\SO(1,1)$ singlets,
\be
\rep{(2,6)}\rightarrow \rep{(2, 1,1)}_{2}+\rep{(2, 1,1)}_{-2}+\rep{(2, 2,2)}_{0}\rightarrow \rep{(2, 2,2)}.
\ee 

An alternative equivalent branching, that is better adapted to the scalar sector is given by 
\begin{eqnarray}
E_{7(7)} &\supset &\SO(6,6)\times \SL (2, \mathds{R})\\ &\supset&
\SO(4,4)\times \SL (2, \mathds{R})\times \SL (2, \mathds{R})_{I}\times \SL (2, %
\mathbb{R})_{II}  \notag \\
&\supset &\SO_{3,3}\times \SL (2, \mathds{R})\times \SL (2, \mathds{R}%
)_{I}\times \SL (2, \mathds{R})_{II}\times \SO(1,1)  \label{0} \\
\mathbf{56} &\rightarrow&\left( \mathbf{12},\mathbf{2}\right) +(\mathbf{32},\mathbf{1}%
)\\
&\rightarrow& \left( \mathbf{8}_{v},\mathbf{2,1,1}\right) +\left( \mathbf{1},\mathbf{%
2,2,2}\right) +\left( \mathbf{8}_{s},\mathbf{1,2,1}\right) +\left( \mathbf{8}%
_{c},\mathbf{1,1,2}\right) ;  \notag \\
&\rightarrow&\left( \mathbf{6},\mathbf{2,1,1}\right) _{0}+\left( \mathbf{1},\mathbf{%
2,1,1}\right) _{2}+\left( \mathbf{1},\mathbf{2,1,1}\right) _{-2}+\left(
\mathbf{1},\mathbf{2,2,2}\right) _{0}  \notag \\
&&+\left( \mathbf{4},\mathbf{1,2,1}\right) _{1}+\left( \mathbf{4}^{\prime },%
\mathbf{1,2,1}\right) _{-1}+\left( \mathbf{4},\mathbf{1,1,2}\right)
_{-1}+\left( \mathbf{4}^{\prime },\mathbf{1,1,2}\right) _{1}.
\end{eqnarray}%
The sequence of maximal subgroups leading to (\ref{0}) can be interpreted as%
\begin{eqnarray}
\Aut\left( \mathfrak{F}\left( \J_{3}^{\mathds{O}_{s}}\right) \right)
&\supset &\Aut\left( \mathfrak{F}\left( \mathds{R}\oplus \J_{2}^{\mathds{O}%
_{s}}\right) \right)  \notag \\
&\supset &\Aut\left( \mathfrak{F}\left( \mathds{R}\oplus \J_{2}^{\mathds{H}%
_{s}}\right) \right) \times \SL(2, \mathds{R})_{I}\times \SL(2, \mathds{R}%
)_{II}  \notag \\
&\supset &\Aut\left( \mathfrak{F}\left( \mathds{R}\oplus \J_{2}^{\mathds{C}%
_{s}}\right) \right) \times \SL(2, \mathds{R})_{I}\times \SL(2, \mathds{R}%
)_{II}\times \SO(1,1),
\end{eqnarray}%
or equivalently as%
\begin{eqnarray}
\Aut\left( \mathfrak{F}\left( \J_{3}^{\mathds{O}_{s}}\right) \right)
&\supset &\Aut\left( \mathfrak{F}\left( \J_{3}^{\mathds{H}_{s}}\right)
\right) \times \SL(2, \mathds{R})  \notag \\
&\supset &\Aut\left( \mathfrak{F}\left( \mathds{R}\oplus \J_{2}^{\mathds{H}%
_{s}}\right) \right) \times \SL(2, \mathds{R})\times \SL(2, \mathds{R})_{II}
\notag \\
&\supset &\Aut\left( \mathfrak{F}\left( \mathds{R}\oplus \J_{2}^{\mathds{C}%
_{s}}\right) \right) \times \SL(2, \mathds{R})_{I}\times \SL(2, \mathds{R}%
)_{II}\times \SO(1,1).
\end{eqnarray}%
By retaining only the $\SL(2, \mathds{R})_{I}\times \SL(2, \mathds{R})_{II}\times \SO(1,1)$ singlets   one obtains%
\begin{equation}
\underset{E_{7(7)}}{\mathbf{56}}\rightarrow \underset{\SO(3,3)\times \SL(2, 
\mathds{R})\times \SL(2, \mathds{R})_{I}\times \SL(2, \mathds{R})_{II}\times
\SO(1,1)}{\left( \mathbf{6},\mathbf{2,1,1}\right) _{0}}\rightarrow \underset{%
\SO(3,3)\times \SL(2, \mathds{R})}{\left( \mathbf{6},\mathbf{2}\right) }.
\end{equation}

For what concerns the maximal compact subgroups, we have 
\begin{eqnarray}
\SU_{8} &\supset &\SO_{6}\times \SO_{6}\times \Un_{1}\\
&\supset &
\SU(2)_{I}\times \SU(2)_{II}\times \SU(2)_{III}\times \SU(2)_{IV}\times \Un(1)\times
\Un(1)_{I}\times \Un(1)_{II}  \notag \\
&\supset &
\SU(2)_{d(I,II)}\times \SU(2)_{d(III,IV)}\times \Un(1)\times
\Un(1)_{I}\times \Un(1)_{II},  \label{1}
\end{eqnarray}%
where $\SU(2)_{d(I,II)}\subset \SU(2)_{I}\times \SU(2)_{II}$ and $%
\SU(2)_{d(III,IV)}\subset \SU(2)_{III}\times \SU(2)_{IV}$ are diagonal embeddings.
 The branchings of (\ref{1}) go as follows:%
\begin{eqnarray}
\mathbf{8} &\rightarrow&
\left( \boldsymbol{4},\mathbf{1}\right) _{1}+\left(
\boldsymbol{1},\mathbf{4}\right) _{-1}\\
&\rightarrow&
\left( \boldsymbol{2},\mathbf{1,1,1}%
\right) _{1,1,0}+\left( \boldsymbol{1},\mathbf{2,1,1}\right)
_{1,-1,0}+\left( \boldsymbol{1},\mathbf{1,2,1}\right) _{-1,0,1}+\left(
\boldsymbol{1},\mathbf{1,1,2}\right) _{-1,0,-1};  \label{2} \\
\mathbf{56} 
&\rightarrow&
\left( \boldsymbol{6},\mathbf{4}\right) _{1}+\left(\boldsymbol{4},\mathbf{6}\right) _{-1}+\left( \overline{\boldsymbol{4}},\mathbf{1}\right) _{3}+\left( \boldsymbol{1},\overline{\mathbf{4}}\right)_{-3}  \notag \\
&\rightarrow&
\left( \boldsymbol{2},\mathbf{2,2,1}\right) _{1,0,1}+\left( \boldsymbol{2}%
,\mathbf{2,1,2}\right) _{1,0,-1}+\left( \boldsymbol{1},\mathbf{1,2,1}\right)
_{1,2,1}+\left( \boldsymbol{1},\mathbf{1,1,2}\right) _{1,2,-1}
  \notag \\
&&
+\left( \boldsymbol{1},\mathbf{1,1,2}\right) _{1,-2,-1} +\left( \boldsymbol{2},\mathbf{1,2,2}\right) _{-1,1,0}+\left( \boldsymbol{1%
},\mathbf{2,2,2}\right) _{-1,-1,0}+\left( \boldsymbol{2},\mathbf{1,1,1}%
\right) _{-1,1,2}
  \notag \\
&&
+\left( \boldsymbol{1},\mathbf{2,1,1}\right)
_{-1,-1,2}+\left( \boldsymbol{2},\mathbf{1,1,1}\right) _{-1,1,-2}+\left(
\boldsymbol{1},\mathbf{2,1,1}\right) _{-1,-1,-2} +\left(
\boldsymbol{1},\mathbf{1,2,1}\right) _{1,-2,1}
  \notag \\
&&
 +\left( \boldsymbol{2},\mathbf{1,1,1}\right) _{3,-1,0}+\left( \boldsymbol{1%
},\mathbf{2,1,1}\right) _{3,1,0}+\left( \boldsymbol{1},\mathbf{1,2,1}\right)
_{-3,0,-1}+\left( \boldsymbol{1},\mathbf{1,1,2}\right) _{-3,0,1}.  \label{3}
\end{eqnarray}%
By retaining only the $\Un(1)_{I}\times \Un(1)_{II}$ singlets  of (\ref{2})
and (\ref{3}) one observes all  gravitini and  gaugini are truncated out, consistent which the fact that bosonic sector has no supersymmetric completion.

For what concerns the scalars, one has%
\begin{eqnarray}
\mathbf{70} &\rightarrow&\left( \boldsymbol{6},\mathbf{6}\right) _{0}+\left(
\boldsymbol{4},\overline{\mathbf{4}}\right) _{-2}+\left( \overline{%
\boldsymbol{4}},\mathbf{4}\right) _{2}+\left( \boldsymbol{1},\mathbf{1}%
\right) _{4}+\left( \boldsymbol{1},\mathbf{1}\right) _{-4}  \notag \\
&\rightarrow&\left( \boldsymbol{2},\mathbf{2,2,2}\right) _{0,0,0}+\left( \boldsymbol{2}%
,\mathbf{2,1,1}\right) _{0,0,2}+\left( \boldsymbol{2},\mathbf{2,1,1}\right)
_{0,0,-2}  \notag \\
&&+\left( \boldsymbol{1},\mathbf{1,2,2}\right) _{0,2,0}+\left( \boldsymbol{1}%
,\mathbf{1,1,1}\right) _{0,2,2}+\left( \boldsymbol{1},\mathbf{1,1,1}\right)
_{0,2,-2}  \notag \\
&&+\left( \boldsymbol{1},\mathbf{1,2,2}\right) _{0,-2,0}+\left( \boldsymbol{1%
},\mathbf{1,1,1}\right) _{0,-2,2}+\left( \boldsymbol{1},\mathbf{1,1,1}%
\right) _{0,-2,-2}  \notag \\
&&+\left( \boldsymbol{2},\mathbf{1,2,1}\right) _{-2,1,-1}+\left( \boldsymbol{%
2},\mathbf{1,1,2}\right) _{-2,1,1}+\left( \boldsymbol{1},\mathbf{2,2,1}%
\right) _{-2,-1,-1}+\left( \boldsymbol{1},\mathbf{2,1,2}\right) _{-2,-1,1}
\notag \\
&&+\left( \boldsymbol{2},\mathbf{1,2,1}\right) _{2,-1,1}+\left( \boldsymbol{2%
},\mathbf{1,1,2}\right) _{2,-1,-1}+\left( \boldsymbol{1},\mathbf{2,2,1}%
\right) _{2,1,1}+\left( \boldsymbol{1},\mathbf{2,1,2}\right) _{2,1,-1}
\notag \\
&&+\left( \boldsymbol{1},\mathbf{1,1,1}\right) _{4,0,0}+\left( \boldsymbol{1}%
,\mathbf{1,1,1}\right) _{-4,0,0}  \notag \\
&\rightarrow&\left( \boldsymbol{3}_{s}+\mathbf{1}_{a},\boldsymbol{3}_{s}+\mathbf{1}%
_{a}\right) _{0,0,0}+\left( \boldsymbol{3}+\mathbf{1},\mathbf{1}\right)
_{0,0,2}+\left( \boldsymbol{3}+\mathbf{1},\mathbf{1}\right) _{0,0,-2}  \notag
\\
&&+\left( \boldsymbol{1},\mathbf{3+1}\right) _{0,2,0}+\left( \mathbf{1},%
\mathbf{1}\right) _{0,2,2}+\left( \mathbf{1},\mathbf{1}\right)
_{0,2,-2}+\left( \boldsymbol{1},\mathbf{3+1}\right) _{0,-2,0}+\left( \mathbf{%
1},\mathbf{1}\right) _{0,-2,2}+\left( \mathbf{1},\mathbf{1}\right) _{0,-2,-2}
\notag \\
&&+\left( \mathbf{2},\mathbf{2}\right) _{-2,1,-1}+\left( \mathbf{2},\mathbf{2%
}\right) _{-2,1,1}+\left( \mathbf{2},\mathbf{2}\right) _{-2,-1,-1}+\left(
\mathbf{2},\mathbf{2}\right) _{-2,-1,1}  \notag \\
&&+\left( \mathbf{2},\mathbf{2}\right) _{2,-1,1}+\left( \mathbf{2},\mathbf{2}%
\right) _{2,-1,-1}+\left( \mathbf{2},\mathbf{2}\right) _{2,1,1}+\left(
\mathbf{2},\mathbf{2}\right) _{2,1,-1}  \notag \\
&&+\left( \mathbf{1},\mathbf{1}\right) _{4,0,0}+\left( \mathbf{1},\mathbf{1}%
\right) _{-4,0,0}.  \label{4}
\end{eqnarray}%
By retaining only the singlets wrt $\Un(1)_{I}\times \Un(1)_{II}$, from (\ref{4})
one obtains%
\begin{equation}
{\mathbf{70}}\rightarrow{\left( \boldsymbol{3}_{s}+\mathbf{1}_{a},%
\boldsymbol{3}_{s}+\mathbf{1}_{a}\right) _{0}+\left( \mathbf{1},\mathbf{1}%
\right) _{4}+\left( \mathbf{1},\mathbf{1}\right) _{-4}}.
\end{equation}%
Therefore, the branching of the scalar fields of $\mathcal{N}=8$, $D=4$
supergravity goes as follows%
\begin{eqnarray}
\frac{E_{7(7)}}{\SU(8)} &\rightarrow &\frac{\SO(3,3)}{\SU(2)_{d(III,IV)}\times
\SU(2)_{d(III,IV)}}\times \frac{\SL(2, \mathds{R})}{\Un(1)}\times \frac{SL(2, 
\mathds{R})_{I}}{\Un(1)_{I}}\times \frac{\SL(2, \mathds{R})_{II}}{\Un(1)_{II}}%
\times \SO(1,1) \notag \\
&\rightarrow &\frac{\SO(3,3)}{\SU(2)_{d(III,IV)}\times \SU(2)_{d(III,IV)}}\times
\frac{\SL(2, \mathds{R})}{\Un(1)},  \label{5}
\end{eqnarray}%
where, from (\ref{4}), the identifications read%
\begin{eqnarray}
T_\mathds{1}\left(\frac{\SL(2, \mathds{R})}{\Un(1)}\times \frac{\SO(3,3)}{\SO(3)\times \SO(3)}\right)
&\sim &\left( \mathbf{1},\mathbf{1}\right) _{4,0,0}+\left( \mathbf{1},%
\mathbf{1}\right) _{-4,0,0}+\left( \boldsymbol{3}_{s},\boldsymbol{3}%
_{s}\right) _{0,0,0};  \label{5-bis} \\
T_\mathds{1}\left(\frac{\SL(2, \mathds{R})_{I}}{\Un(1)_{I}}\right)  &\sim &\left( \boldsymbol{1},\mathbf{1%
}\right) _{0,2,0}+\left( \boldsymbol{1},\mathbf{1}\right) _{0,-2,0}; \\
T_\mathds{1}\left(\frac{\SL(2, \mathds{R})_{II}}{\Un(1)_{II}} \right) &\sim &\left( \boldsymbol{1},%
\mathbf{1}\right) _{0,0,2}+\left( \boldsymbol{1},\mathbf{1}\right) _{0,0,-2};
\\
T_\mathds{1}\left(\SO(1,1)\right) &\sim &\left( \boldsymbol{1}_{a},\boldsymbol{1}_{a}\right) _{0,0,0},
\end{eqnarray}%
where the covariance is $\SU(2)_{d(I,II)}\times \SU(2)_{d(III,IV)}\times
\Un(1)\times \Un(1)_{I}\times \Un(1)_{II}$. Note that $\left( \boldsymbol{3}_{s},%
\mathbf{1}_{a}\right) _{0,0,0}+\left( \mathbf{1}_{a},\boldsymbol{3}%
_{s}\right) _{0,0,0}$, despite being singlets under $\Un(1)_{I}\times \Un(1)_{II}$,
do not correspond to any (reductive) coset. Since 
\be
\Aut\left( \mathfrak{F}\left( \mathds{R}\oplus \J_{2}^{\mathds{C}%
_{s}}\right) \right) \cong \SL(2, \mathds{R})\times \SO(3,3),
\ee
 the coset of
the $\mathcal{N}=0$, $D=4$ Maxwell-Einstein theory based on $\mathds{R}%
\oplus \J_{2}^{\mathds{C}_{s}}$ can be identified with (\ref{5}), or
equivalently with (\ref{5-bis}).

Apply the charge quantisation condition, we have automorphism group $\SL(2, \mathds{Z})\times \SL(4, \Z)$, which will be related to linear basis changes of ideals in the higher composition law. The quantised black hole charges $\cq=a_{A[ij]}\ket{A[ij]}$, $i,j=1,2,3,4$ belong to
\be
\FTS(\Z\oplus\Gamma_{1,5}(\Z))\cong \Z^2\otimes \wedge^2 \Z^4,
\ee
which one can regard as the space of pairs of integral 2-forms,
\be
\cq=\begin{pmatrix}a_{0[ij]}\\a_{1[ij]}\end{pmatrix}.
\ee

From the embedding of the $STU$ model, there is a natural linear map of the black hole charges 
\be
\text{id}\otimes \wedge_{2,2}: \Z^2\otimes\Z^2\otimes\Z^2\rightarrow \Z^2\otimes \wedge^2 \Z^4,
\ee
which in terms of the hypermatrix is given by 
\be
a_{ABC}\mapsto a_{A[ij]}= \begin{pmatrix}0 & a_{ABC}\\-a_{ACB}&0\end{pmatrix}.
\ee
In order to state the bijection between the U-duality equivalence classes of black holes and quadratic ideal classes we need a few more notions from the theory of higher rank ideals. A rank $n$ ideal of $S$ is an $S$-submodule of $K^n$, where $K=S\otimes\mathds{Q}$, of rank $2n$ as a $\Z$-module.  The determinant of a rank $n$ ideal $M$ is denoted $\Det(M)$ and defined as the ideal in $S$ generated by all $\det(\mathbf{M})$, where $\mathbf{M}\in M^n$ is regarded as an $n\times n$ matrix.  A $k$-tuple of oriented ideals $(M_1, \ldots, M_k)$ of ranks $n_1,\ldots n_k$ is said to be balanced if $\prod_{i=1}^{k}\Det(M_i)\subseteq S$ and $\prod_{i=1}^{k}N(M_i)=1$.

With the notion of balanced $k$-tuples of rank $n$ ideals in hand, we can state the parametrisation of U-duality equivalence classes of black hole solutions in terms of ideal classes in quadratic orders:

 \begin{theorem}[Parametrisation of equivalence classes of pairs of alternating forms \cite{Bhargava:2004}]
There is a canonical bijection between the set of nondegenerate $\SL(2, \Z)\times \SL(4, \Z)$-orbits on the space $\Z^2\otimes \wedge^2 \Z^4$  of quartic norm $\Delta(\cq)$, and the set of equivalence classes of pairs $(S, (I, M))$, where $S$ is a nondegenerate oriented quadratic ring with discriminant $D=-\Delta(\cq)$, and $(I, M)$ is a balanced doublet of ideals of ranks one and two, respectively. 
\end{theorem}
Under the mapping embedding the $STU$ model into the $\SL(2, \Z)\times \SL(4, \Z)$ Einstein-Maxwell-Scalar theory, the equivalence classes of pairs $(S, (I_S, I_T, I_U))$ get mapped into the equivalence classes of pairs of the form $(S, (I_S, I_T\oplus I_U))$; the embedding corresponds to the fusion of rank one ideals $(I_T, I_U)$ into a single rank two ideal $(I_T\oplus I_U)$. Clearly, every $\SL(2, \Z)\times \SL(2, \Z)\times \SL(2, \Z)$-equivalence class maps into an $\SL(2, \Z)\times \SL(4, \Z)$-equivalence class. What is less obvious is that every $\SL(2, \Z)\times \SL(4, \Z)$-equivalence class has a representative in the image of $\text{id}\otimes \wedge_{2,2}$. Elements of $\Z^2\otimes \wedge^2 \Z^4$ can be ``diagonalised'' by $\SL(2, \Z)\times \SL(4, \Z)$ to lie in a subspace isomorphic to $\Z^2\otimes\Z^2\otimes\Z^2$. This follows directly from the fact that  any torsion-free module over $S$ is a direct sum of rank one ideals. Thus, the embedding map $(S, (I_S, I_T, I_U))\mapsto (S, (I_S, I_T\oplus I_U))$ is surjective at the level of equivalence classes and we conclude that every $\cq\in\Z^2\otimes \wedge^2 \Z^4$ can be ``diagonalised'' to an element of the form $\text{id}\otimes\wedge_{2,2}(a_{ABC})$.  

This allows one to use projectivity  of elements in   $ \Z^2\otimes\Z^2\otimes\Z^2$ to define projectivity of elements in  $\Z^2\otimes \wedge^2 \Z^4$: $\cq\in\Z^2\otimes \wedge^2 \Z^4$ is projective if and only if it is $\SL(2, \Z)\times \SL(4, \Z)$-equivalent to $\text{id}\otimes \wedge_{2,2}(a_{ABC})$ for some projective $a_{ABC}$. As before, restricting to projective elements yields a group law:

 \begin{theorem}[Group law on projective pairs integral   2-forms \cite{Bhargava:2004}]\label{gl24} For all $D=0,1 \mod 4$,  the set of ~$\SL(2,\Z)\times\SL(4, \Z)$-equivalence classes of projective integral 2-forms, $\cq\in\Z^2\otimes \wedge^2\Z^4$ with fixed quartic norm $D=-\Delta(\cq)$ is a unique group $\text{\emph{Cl}}(\Z^2\otimes \wedge^2 \Z^4 ;D)$ such that $[a_{ABC}]\mapsto [\text{id}\otimes \wedge_{2,2}(a_{ABC})]$ is group homomorphism from $\text{\emph{Cl}}(\Z^2\otimes \Z^2\otimes\Z^2; D)$.
\end{theorem}

One can also maps to integral binary quadratic forms $\Z^2\otimes \wedge^2\Z^4\rightarrow \sym^2(\Z^2)^*$ using 
\be
f^\cq(x,y) = x^A a_{A[ij]}a_{A'[i'j']}\varepsilon_{iji'j'}x^{A'}.
\ee
The discriminant $D$ of $f^\cq$ is the given by $D=-\Delta(\cq)$. Restricting to projective $\cq$ implies primitive $f^\cq$. Returning to the S-module point of view, projective implies the 2-tuples $(I, M)$ satisfy $I\Det(M)=S$, which implies $(S, (I, M))\cong (S, (I, S\oplus I^{-1}))$. Hence, we have a group isomorphism $\text{{Cl}}(\Z^2\otimes \wedge^2 \Z^4 ;D)\rightarrow \text{{Cl}}(\sym^2(\Z^2)^* ;D)\cong \text{Cl}^+(S(D))$. The projective black hole solutions are in one-to-one correspondence with the elements of the narrow class group.

\subsubsection{The $\SL(6, \Z)$ Einstein-Maxwell-Scalar theory}\label{sl6}

The final example of Barghava's higher composition laws associated to quadratic orders corresponds to case 6' of \autoref{tab:FTSsummary} over the integral split complexes $\mathfrak{C}_s$,   $\FTS_{\SL_6(\Z)}:=\FTS(\J_{3}^{\mathfrak{
C}_s})$, where $\J_{3}^{\mathfrak{C}_s}$ is the set of $3\times 3$ Hermitian matrices over $\mathfrak{C}_s$.   The automorphism group is given by 
\be
\Aut(\FTS(\J_{3}^{\mathfrak{C}_s}))  \cong  \SL(6, \Z).
\ee
It corresponding theory of type $E_7$ is given by   Einstein-Hilbert gravity coupled to 10 Abelian gauge fields, which together with their duals belong to the $\rep{20}\cong \wedge^3(\R^6)$ of $\SL(6, \R)$, and 20  real scalar fields parametrising the coset,
\be
\frac{\SL(6, \R)}{\SO(6)}.
\ee
It has been treated previously in \cite{Marrani:2017aqc, Marrani:2019jvd}. Although it does not admit a supersymmetric completion\footnote{$\N=4$ is ruled out by the even number of scalars, $\N=3$ is ruled out by $20\not=6n$, $\N=2$ is ruled out by the requirement that the scalar manifold be special K\"ahler. This leaves the final possibility of $\N=1$ supergravity coupled to ten vector multiplets and 10 chiral multiplets, but this is ruled out by the fact that the scalar manifold of chiral multiplets must be K\"ahler, as well as from the fact that the kinetic vector matrix must be holomorphic. } in $D=1+3$, as well as a theory of type $E_7$ in its own right it can be regarded as a consistent truncation of $\N=8$ supergravity effected through the branching under 
\be
\SL(6, \R)\subset \SL(2, \R)\times[ \SO(1,1)\times \SL(6, \R)]\subset  \SL(2, \R)\times \SO(6,6, \R)\subset E_{7(7)},
\ee
so that the $28+28$ vectors  and their duals break as 
\be
\rep{56}\rightarrow \rep{(2,12)+(1,32)}\rightarrow\rep{6}_{-2}+\rep{6}'_{-2}+\rep{20}_{0}\rightarrow \rep{20}
\ee
where we retain only the $\SL(2,\R)$ and $\SO(1,1)$ singlets at the first and second branchings, respectively.   From the above we see that the $\SL(2, \Z)\times \SL(4, \Z)$ theory may be embedded  in the $\SL(6, \Z)$ Einstein-Maxwell-scalar theory. 

The quantized black hole charges $\cq=a_{[abc]}\ket{[abc]}$, $a,b,c=1,\ldots6,$ belong to
\be
\FTS(\J_{3}^{\mathfrak{C}_s})\cong \wedge^3 \Z^6.
\ee

From the embedding of the $STU$ model, there is a natural linear map of the black hole charges 
\be\label{STU-SL6}
\wedge_{2,2,2}: \Z^2\otimes\Z^2\otimes\Z^2\hookrightarrow \wedge^3 \Z^6,
\ee
which in terms of the canonical FTS basis  is given by 
\be
\begin{array}{llllll}
\wedge_{2,2,2}:& \Z\oplus \Z\oplus\mathfrak{J}_{STU}\oplus\mathfrak{J}_{STU}& \hookrightarrow &\Z\oplus \Z\oplus\J_{3}^{\mathfrak{C}_s}\oplus\J_{3}^{\mathfrak{C}_s}\\[5pt]
&(\alpha, \beta, (A_1, A_2, A_3), (B_1, B_2, B_3))& \mapsto &(\alpha, \beta, \diag(A_1, A_2, A_3), \diag(B_1, B_2, B_3)).
\end{array}
\ee
As before, the FTS quartic norm can be regarded as the discriminant, $\Delta(\mathcal{Q})=-D$.    

 \begin{theorem}[Parametrisation of $\SL(6, \R)$-equivalence classes in $\wedge^3(\Z^6)$ \cite{Bhargava:2004}]
There is a canonical bijection between the set of nondegenerate $\SL(6, \Z)$-orbits on the space $\wedge^3 \Z^6$  of quartic norm $\Delta(\cq)$, and the set of isomorphism classes of pairs $(S,  M)$, where $S$ is a nondegenerate oriented quadratic ring with discriminant $D=-\Delta(\cq)$, and $M$ is an equivalence class of  balanced  of ideals of rank three. 
\end{theorem}

Once again the proof of this statement hinges  on the identification of $\cq\in\wedge^3\Z^6$ with $a_{[ijk]}$ in 
\be
\det(\alpha_i, \alpha_j, \alpha_k) = c_{[ijk]}+a_{[ijk]}\tau,
\ee
where $\{\alpha_i\}_{i=1}^{6}$  is an oriented $\Z$-basis for $M$. The rather complicated looking unique solution for the $c_{[ijk]}$ is again quite simply given by the triple product
\be
c_{[ijk]} = -\frac{1}{2}\left(T(a)_{[ijk]} +(\Det a\mod 4) a_{[ijk]}\right),
\ee
emphasising the role played by the FTS. 

Under the mapping embedding the $STU$ model into the $\SL(6, \Z)$ Einstein-Maxwell-Scalar theory, the equivalence classes of pairs $(S, (I_S, I_T, I_U))$ get mapped into the equivalence classes of pairs of the form $(S, (I_S\oplus I_T\oplus I_U))$; the embedding corresponds to the fusion of triples of rank one ideals $((I_S, I_T, I_U)$ into a single rank three idea. Every $\SL(2, \Z)\times \SL(2, \Z)\times \SL(2, \Z)$-equivalence class maps into an $\SL(6, \Z)$-equivalence class. Moreover, every $\SL(6, \Z)$-equivalence class has a representative in the image of $\wedge_{2,2,2}$. Elements of $ \wedge^3 \Z^6$ can be ``diagonalised'' by $\SL(6, \Z)$ to lie in a subspace isomorphic to $\Z^2\otimes\Z^2\otimes\Z^2$.  Thus, the embedding map $(S, (I_S, I_T, I_U))\mapsto (S, (I_S\oplus I_T\oplus I_U))$ is surjective at the level of equivalence classes and we conclude that every $\cq\in \wedge^3 \Z^6$ can be ``diagonalised'' to an element of the form $\wedge_{2,2,2}(a_{ABC})$.  

As before, this allows one to use projectivity  of elements in   $ \Z^2\otimes\Z^2\otimes\Z^2$ to define projectivity of elements in  $ \wedge^3 \Z^6$. Restricting to projective elements the set of equivalence classes is endowed with  a (trivial) group structure:

 \begin{theorem}[Group law on projective  integral   3-forms \cite{Bhargava:2004}]\label{gl24} For all $D=0,1 \mod 4$,  the set of ~$\SL(6, \Z)$-equivalence classes of projective integral 3-forms, $\cq\in  \wedge^3\Z^6$ with fixed quartic norm $D=-\Delta(\cq)$ is a unique group $\text{\emph{Cl}}(\wedge^3 \Z^4 ;D)$ such that $[a_{ABC}]\mapsto [\wedge_{2,2,2}(a_{ABC})]$ is group homomorphism from $\text{\emph{Cl}}(\Z^2\otimes \Z^2\otimes\Z^2; D)$ and:
 \begin{enumerate}
 \item $\SL(6, \Z)$ is transitive on the set of projective elements with fixed $D=-\Delta(\cq)$. Hence, $\text{\emph{Cl}}(\wedge^3 \Z^6 ;D)$ is a one-element group. 
 \item An integer is a fundamental discriminant if it is square-free and either $1\mod 4$ or $4m$, where $m=2,3 \mod 4$. If $D=-\Delta(\cq)$  is a fundamental discriminant then $\cq$ is projective. Hence, up to $\SL(6, \Z)$-equivalence there is a unique $Q\in \wedge^3 \Z^4$ with $D=-\Delta(\cq)$  a fundamental discriminant. 
 \end{enumerate}
\end{theorem}

We conclude, as observed in \cite{Borsten:2010aa, Borsten:2010ths}, that there is up to U-duality equivalence a unique projective extremal black hole solution for fixed Bekenstein-Hawking entropy and a unique extremal black solution for $(\frac{1}{\pi}S_{\text{BH}})^2$ a fundamental discriminant.

\subsubsection{$\mathcal{N}=8$ supergravity and the $\SO(6,6)$ Einstein-Maxwell-scalar theory}

A particularly important example of   extremal black hole solutions are those of  $\N=8$ supergravity, the low-energy effective field theory limit of type IIA/B string (M-theory) on a 6-torus (7-torus). In this case the black hole charges are elements of $\FTS_{\N=8}=\FTS(\J^{\mathfrak{O}_s}_{3})$.  Rather than the $A_n$ type automorphism (electromagnetic duality)  groups appearing in all the previous examples, here it  is given by the exceptional $E_{7(7)}(\Z)$ \cite{Cremmer:1979up, Hull:1994ys}. This rather obscures any potential connection to ideal classes in quadratic orders, where previously an electromagnetic duality  transformation corresponded to an arbitrary orientation preserving  basis changes of the form $\SL(n, \Z)\times\SL(n, \Z)\times \cdots$. Bhargava does not provide a  composition law for this exceptional case, likely for this very reason; it is not straightforward to identify suitable ideals in a quadratic order corresponding to the $\rep{56}$ of $E_{7(7)}(\Z)$. Nonetheless, the natural inclusions
\be
\FTS(\Z)\hookrightarrow \FTS(\Z\oplus \Z)\hookrightarrow \FTS(\Z\oplus \Z\oplus\Z)\hookrightarrow \FTS(\Z\oplus \Gamma_{5,1}(\Z))\hookrightarrow\FTS(\J^{\mathfrak{C}_s}_{3})\hookrightarrow \FTS(\J^{\mathfrak{H}_s}_{3})\hookrightarrow \FTS(\J^{\mathfrak{O}_s}_{3})
\ee
and their associated groups of type $E_7$,  
\be
\SL(2, \Z) \subset [\SL(2, \Z)]^2 \subset[\SL(2, \Z)]^3\subset  \SL(2, \Z) \times \SL(4, \Z) \subset  \SL(6, \Z) \subset  \SO(6, 6; \Z) \subset  E_{7(7)}(\Z),
\ee 
allows for the same  definition of projectivity \cite{Krutelevich:2004} to be used for $\FTS(\J^{\mathfrak{O}_s}_{3})$:

 \begin{definition} An element in $\FTS(\J^{\mathfrak{O}_s}_{3})$ is projective if it is $E_{7(7)}(\Z)$-equivalent to an element lying in the image of projective elements in $\Z^2\otimes \Z^2\otimes \Z^2$ under  the inclusion map given by 
 \be\label{STU-N8}
\begin{array}{llllll}
\imath_{E_7}:& \Z\oplus \Z\oplus\mathfrak{J}_{STU}\oplus\mathfrak{J}_{STU}& \hookrightarrow &\Z\oplus \Z\oplus\J_{3}^{\mathfrak{O}_s}\oplus\J_{3}^{\mathfrak{O}_s}\\[5pt]
&(\alpha, \beta, (A_1, A_2, A_3), (B_1, B_2, B_3))& \mapsto &(\alpha, \beta, \diag(A_1, A_2, A_3), \diag(B_1, B_2, B_3)).
\end{array}
\ee
\end{definition}
This is the same inclusion and  definition  used by Bhargava in the previous example of integral 3-forms. In that case, Bhargava showed that every integral 3-form is  $\SL(6, \Z)$-equivalent to some element in the image of the inclusion \eqref{STU-SL6} through the bijection   to 
isomorphism classes of pairs $(S, M)$. In the absence of  an equivalent  bijection, Krutelevich explicitly  demonstrated that every  element in $\FTS(\J^{\mathfrak{O}_s}_{3})$ is $E_{7(7)}(\Z)$-equivalent to an element lying in the image of \eqref{STU-N8}. Note, Krutelevich's proof relied crucially on the fact that every $3\times 3$ Hermitian matrix defined over the \emph{split} integral octonions can be diagonalised \cite{Krutelevich:2002}. By contrast, \emph{not} every $3\times 3$ Hermitian matrix defined over Coxeter's ring  of  integral \emph{division}  octonions can be diagonalised \cite{Elkies:1996}; the existence of zero-divisors was essential in the proof of the former case. This has a bearing on the applicability of Bhargava's higher composition laws to the magic supergravities proposed in \cite{Gunaydin:2019xxl}, as discussed in more detail in \autoref{comments}. 

Using this notion of projectivity and the orbits classification of \cite{Krutelevich:2004} we have the  trivial group law directly analogous to Bhargava's $\wedge^3\Z^6$ example:

 \begin{theorem}[Group law on projective  $E_{7(7)}(\Z)$-equivalence classes] For all $D=0,1 \mod 4$,  the set of ~$E_{7(7)}(\Z)$-equivalence classes of projective  $\cq\in  \FTS(\J^{\mathfrak{O}_s}_{3})$ with fixed quartic norm $D=-\Delta(\cq)$ is a unique group $\text{\emph{Cl}}(\FTS(\J^{\mathfrak{O}_s}_{3}) ;D)$ such that $[a_{ABC}]\mapsto [\imath_{E_7}(a_{ABC})]$ is group homomorphism from $\text{\emph{Cl}}(\Z^2\otimes \Z^2\otimes\Z^2; D)$ and:
 \begin{enumerate}
 \item $E_{7(7)}(\Z)$ is transitive on the set of projective elements with fixed $D=-\Delta(\cq)$. Hence, $\text{\emph{Cl}}(\FTS(\J^{\mathfrak{O}_s}_{3}) ;D)$ is a one-element group. 
 \item If $D=-\Delta(\cq)$  is a fundamental discriminant then $\cq$ is projective. Hence, up to $E_{7(7)}(\Z)$-equivalence there is a unique $Q\in\FTS(\J^{\mathfrak{O}_s}_{3})$ with $D=-\Delta(\cq)$  a fundamental discriminant. 
 \end{enumerate}
\end{theorem}

All extremal black holes with fixed Bekenstein-Hawking entropy in $\N=8$ supergravity with projective charge configurations are U-duality related  \cite{Borsten:2009zy, Borsten:2010aa}. This is not true for non-projective black holes; the discrete invariants \eqref{Zinvariants} can be used to demonstrate that there are charge configurations with the same Bekenstein-Hawking entropy that are not U-duality related \cite{Borsten:2009zy, Borsten:2010aa}. Ideally, we would like a parametrization result for generic charge configurations.  Note that a precisely analogous analysis applies to the $\SO(6,6)$ Einstein-Maxwell-scalar theory of type $E_{7}$ given by the FTS $\FTS(J_{3}^{\Q_s})$, which has 16 gauge potentials and 36 scalars parametrising $\SO(6,6)/[\SO(6)\times\SO(6)]$. In particular, we can use the same inclusion map to define projectivity and we have the same trivial group law on projective equivalence classes. A complete description  of this case  is given in \cite{Borsten:2010aa}. 
\subsection{Comment on the magic  $\mathcal{N}=2$ supergravity theories}\label{comments}

In the pioneering work \cite{Gunaydin:1983bi, Gunaydin:1983rk, Gunaydin:1984ak} it was shown that $D=5, \N=2$ supergravity theories with  symmetric scalar manifolds are intimately related to Euclidean cubic Jordan algebras. In particular, if the Abelian gauge fields are to transform irreducibly under the global symmetry group of the Lagrangian, then there are four possibilities completely determined by the choice of cubic Jordan algebra $\J_{3}^{\da}$, where $\da$ is one of the normed division algebras $\R, \C, \Q, \Oct$. These are typically referred to as the magic supergravities since their global symmetries form a row of the Freudenthal-Rosenfeld-Tits magic square of Lie algebras. There is a further countable family   given by the $(1,n)$-spin-factor Jordan algebras $\R\oplus \Gamma_{1,n}$, which are sometimes referred to as the generic Jordan supergravities. 

Dimensionally reducing the magic supergravities  on a circle one obtains the $D=4$  magic/generic Jordan supergravities with electromagnetic duality groups given by the automorphism groups of the associated FTS, $\FTS(\J)$, over the corresponding $D=5$ Jordan algebra $\J=\J_{3}^{\da}$ or $\R\oplus \Gamma_{1,n}$. Hence, in all cases they are theories of type $E_7$ with electromagnetic duality groups given in \autoref{tab:FTSsummary}. Given the relationship between theories of type $E_7$ and higher composition laws already presented, it would perhaps be natural to expect that the magic and generic Jordan supergravities should correspond to Bhargava's composition laws. Indeed, it was suggested in \cite{Gunaydin:2019xxl} that the magic complex theory should be related to the composition law on integral 3-forms discussed  in \autoref{sl6}. Here  we discuss why this case, and other related examples, are  not directly related to the specific composition laws introduced by Bhargava. That is not to say, however, there is \emph{no} composition law and/order ideal classes associated to such theories.

Let us begin with the magic complex case suggested in \cite{Gunaydin:2019xxl}. As observed in \cite{Krutelevich:2004, Borsten:2010aa, Borsten:2010ths} and developed in \autoref{sl6}, the space of real 3-forms $\wedge^3(\R^6)$ can be identified with $\FTS(\J_{3}^{\C_s})$ with automorphism group  $\SL(6, \R)$. The corresponding theory of type $E_7$ was the Maxwell-Einstein supergravity of \autoref{sl6}, which does not admit a supersymmetric extension. Switching from $\J_{3}^{\C_s}$ to $\J_{3}^{\C}$ gives the magic complex $\N=2$ supergavity once the fermionic sector is included. The change from $\C_s$ to $\C$ implies a different real form for the electromagnetic duality group: $\SL(6, \R)\rightarrow \SU(3,3)$. This already presents the first obstacle. The correspondence to the higher composition required that the discrete subgroup, obtained by imposing the Dirac-Zwanziger-Schwinger  quantisation, corresponded to linear basis changes of the associated  rank-3 ideals, that is $\SL(6, \Z)$ and not $\SU(3, 3; \Z)$.  Already over the reals $\SL(6, \R)$ and  $\SU(3,3)$ have distinct orbits structures\footnote{For $\Delta>0$ elements the maximally split real form $\SL(6, \R)$ has a unique orbit, while $\SU(3, 3)$ has two orbits \cite{Krutelevich:2004, Bellucci:2006xz, Marrani:2019jvd}.} on the rank four elements of $\FTS(\J_{3}^{\C_s})$ and $\FTS(\J_{3}^{\C})$, so the connection to Bhargava's composition law on 3-forms is lost. 

At this stage one might object that the $\N=8$ case $E_{7(7)}(\Z)$ also lacked an interpretation as the set of basis changes of some ideal classes. One could however ``diagonalise'' the non-degenerate elements of $\FTS(\J_{3}^{\Oct_s})$ and so apply the notion of projectivity.  Perhaps, then, there is some hope for the  octonionic magic $\N=2$ supergravity corresponding $\FTS(\J_{3}^{\Oct})$  with automorphism group $E_{7(-25)}(\Z)$.  However, we immediately run into a problem. The diagonalisation proof of \cite{Krutelevich:2004} relied on the diagonisability of all elements in $\J_{3}^{\mathfrak{O}_s}$ \cite{Krutelevich:2002}, but it is known that there are elements in $\J_{3}^{\mathfrak{O}}$ that cannot be diagonalised \cite{Elkies:1996}  and hence the notion of projectivity cannot be applied, at least not without significant further work. 

\subsection{Comments on the relations to $D=0+3$} 

A powerful technique in the construction stationary black solutions is the time-like dimensional reduction to $D=0+3$, as pioneered in \cite{Breitenlohner:1987dg}. On performing the dimensional reduction and dualizing all resulting vectors to scalars, the U-duality is enhanced $G_4\rightarrow G_3$. The scalars parametrise a non-compact coset $G_3/H_3$, where $H_3$ is a non-compact real form of the maximal compact subgroup of $G_3$. For example, in the case of $D=1+3$ $\N=8$ supergravity we have 
\be
E_{7(7)} \longrightarrow  E_{8(8)}, \qquad \frac{E_{7(7)}}{\SU(8)}\longrightarrow  \frac{E_{8(8)}}{\SO^\star(16)}.
\ee
The nilpotent orbits of $G_3$ can then be used to classify the classical stationary black hole solutions in $D=1+3$ \cite{Gunaydin:2007bg,Bergshoeff:2008be, Bossard:2008sw, Bossard:2009we,Bossard:2009at}\footnote{In the case of the $STU$ model this can also be applied to classify the entanglement of four-qubit states \cite{Borsten:2010db}}. For the six theories of type $E_7$ related to quadratic higher composition laws we have, under time-like dimensional reduction, the following U-duality groups \cite{Breitenlohner:1987dg}:
\be
\begin{array}{llllllllllllllll}
 \SL(6, \Z)&\longrightarrow & E_{6(6)}(\Z)\\
 \SL(2, \Z)\times\SL(4, \Z) &\longrightarrow & \SO(5,5; \Z) \\
 \SL(2, \Z)\times\SL(2, \Z)\times\SL(2, \Z) &\longrightarrow & \SO(4,4; \Z)\\
 \SL(2, \Z)\times\SL(2, \Z) &\longrightarrow & \SO(3,4; \Z)\\
  \SL(2, \Z) &\longrightarrow & G_{2(2)}(\Z).
\end{array}
\ee
Amazingly, Bhargava independently identified  these very groups by considering fusings of the ideals entering the parametrization of the orbits in terms of quadratic orders. Using these principles Bhargava was able to identify cubic composition laws, which also have corresponding black $p$-branes.  The dimensional reduction of (super)gravity theories knows about higher composition laws and vice versa! Note, the nilpotent orbits of $G_3$ enter into the scattering amplitudes and BPS instantons of closed superstring theory, see for example \cite{Green:2011vz}, and so we should expect a connection to Bhargava there too. We leave this for future work. 

\section{Conclusions} We have reviewed the relationships between black hole charge orbits and the higher composition laws of Bhargava that were introduced in \cite{Borsten:2008zz, Borsten:2008wd, Borsten:2009zy, Borsten:2010aa, Borsten:2010ths}, including two new examples. We gave a more complete description of how this works and how the black hole charge orbits are parametrised by ideal classes through Bhargava's work. We have explained why this correspondence should exist; it comes down to the dual role of prehomogenous vector spaces in constructing higher composition laws and  theories of type $E_7$, which include all the (super)gravity theories of relevance here. In particular, we see the triple product of groups of type $E_7$ appear in the bijection between black hole charge orbits and ideal classes. Finally, we have noted the relationship between higher composition laws and dimensional reduction. We also claimed that all 14 of Bhargava's higher composition laws are related to black $p$-branes in (super)gravity theories. For example, using the analysis of \cite{Ferrara:2010ug} we see that the two-centered black hole/black string solutions of the $\mathcal{N}=2$, $D=1+4$
magic model based on $\J_{3}^{\R}$ correspond to case I.(8) of the list of regular prehomogeneous vector spaces given in \S 7 of \cite{sato_kimura_1977} and Bhargava's composition law on the $\text{GL}(2, \Z)\times \SL(3, \Z)$-equivalence classes of $\Z^2\otimes \sym^2\Z^3$, which are related to order two ideal classes in cubic rings \cite{bhargava2004higher}. We leave the details of this for future work.

\section*{Note added} During the  completion of this manuscript the preprint \cite{Banerjee:2020lxj} appeared. It has significant overlap with our discussion of the $STU$ model, but also contains  interesting and significant developments regarding the $STU$ charge orbits not contained here.

\section*{Further note added} In a revised version of \cite{Banerjee:2020lxj} the authors state ``After submitting the paper to the arXiv, we became aware of Refs. \cite{Borsten:2008zz,  Borsten:2010ths} where possible connection between Bhargava's cube and the black hole charges in the STU model was discussed. Although the original references are not available in the arXiv, a review of these results can be found in the recent arXiv paper [the present article].''

We therefore make clear that the identification of the $STU$ black hole charge cube \cite{Duff:1995sm} and Bhargava's  cube \cite{Bhargava:2004} was made explicitly in  \cite{Borsten:2007rtn, Borsten:2008zz, Borsten:2008wd,  Borsten:2010ths}. The application of Bhargava's cube higher composition law to the $STU$ charge orbit classification was made  in \cite{Borsten:2010ths}, with further relations discussed in the five  references \cite{Borsten:2007rtn,Borsten:2008wd, Borsten:2009zy, Borsten:2010aa, Borsten:2010ths}.  

\section*{Acknowledgments}

 LB would like to thank Atish Dabholkar and Dan Waldram for useful conversations. The work of LB has been supported by a Schr\"odinger Fellowship and the Leverhulme Trust. 
MJD is grateful to  Philip Candelas for hospitality at the Mathematical Institute, University of Oxford, to Marlan Scully for his hospitality in the Institute for Quantum Science and Engineering, Texas A\&M University, and to the Hagler Institute for Advanced Study at Texas A\&M for a Faculty Fellowship. The  work of MJD was supported in part by the STFC under rolling grant ST/P000762/1.

%\bibliography{Ref_Library}
%\bibliographystyle{utphys}

\providecommand{\href}[2]{#2}\begingroup\raggedright\endgroup

\end{document}